\documentclass[aps,final,reprint,twocolumn,floatfix,superscriptaddress,notitlepage]{revtex4-2}
\usepackage{graphicx,amssymb,soul,color,amsmath,bm}
\usepackage{epstopdf,hyperref}
\usepackage{braket,dsfont,nicefrac}
\usepackage[mathcal]{euscript}
 \usepackage{subcaption}
\usepackage{ragged2e}
\usepackage{etoolbox}
\usepackage{ulem}
\makeatletter \long\def\@makecaption#1#2{  \par  \begingroup    \small    \justifying    \textbf{#1.} #2\par  \endgroup}

\hypersetup{colorlinks=true,citecolor=blue,linkcolor=magenta}

\sethlcolor{yellow}
\allowdisplaybreaks

\usepackage{xcolor}

\begin{document}

\title{Dispersion of low-energy optical excitations in NiPS$_3$ explained by singlet polaron theory} 

\author{I. J. Hamad}
\affiliation{Instituto de Física Rosario (CONICET) and Facultad de Ciencias Exactas, Ingeniería y Agrimensura, Universidad Nacional de Rosario, 2000 Rosario, Argentina}

\author{C. S. Helman}
\affiliation{Instituto de Nanociencia y Nanotecnolog\'{\i}a CNEA-CONICET, GAIDI, Centro At\'{o}mico Bariloche and Instituto Balseiro, 8400 Bariloche, Argentina}

\author{A. E. Feiguin}
\affiliation{Physics Department, Northeastern University, Boston, MA 02115, USA}

\author{L. O. Manuel}
\affiliation{Instituto de Física Rosario (CONICET) and Facultad de Ciencias Exactas, Ingeniería y Agrimensura, Universidad Nacional de Rosario, 2000 Rosario, Argentina}

\author{A. A. Aligia}
\affiliation{Instituto de Nanociencia y Nanotecnolog\'{\i}a CNEA-CONICET, Centro At\'{o}mico Bariloche and Instituto Balseiro, 8400 Bariloche, Argentina}

\begin{abstract}
Optical and resonant inelastic X-ray scattering (RIXS) experiments in the van der Waals antiferromagnet NiPS$_3$ have revealed narrow excitonic resonances near 1.5 eV whose microscopic origin remains under debate. In this work we study the momentum dependence of these excitations using the singlet-polaron theory (SPT) previously developed for NiPS$_3$. Within its framework, we calculate the spectral function of the corresponding effective model and extract the dispersion of the resulting quasiparticles.
The theory naturally predicts two singlet excitations whose dynamics arise from their propagation in the zig-zag antiferromagnetic background ($S=1$) coupled to two-magnon processes. We show that the dispersion of the lowest-energy quasiparticle agrees very well with the excitonic dispersion measured by RIXS, reproducing both its overall shape and bandwidth along high-symmetry directions of the Brillouin zone. Our results provide a consistent interpretation of the optical excitations in terms of dispersive singlet polarons, and suggest that similar mechanisms may govern excitonic dynamics in other layered van der Waals antiferromagnets.

\end{abstract}

\maketitle

\section{Introduction}
\label{intro}
The recent identification of atomically thin, exfoliated van der Waals magnets ~\cite{Kuo2016,Lee2016,Burch2018,Gibertini2019}, especially members of the MPX$_3$ family (M = Mn, Fe, Ni) ~\cite{Joy1992,Wildes2015,Kim2018,Kim2019,Kang2020,Hwangbo2021,Olsen2021,Kim2021,Wang2021,Ho2021,Belvin2021,Afanasiev2021,Wang22,Wildes2022,Mi2022,Mehlawat2022,Dirnberger2022,Scheie2023,plumley2023,DiScala2023,Jana2023,Klaproth2023,KimKim2023,Kim2023,He2024,Allington2025,Wars25,Wars25b,Wang25,Allington2025,Schue26,Cojo26,Rybak26}, has stimulated a great deal of experimental and theoretical work. Compounds in this class typically develop long-range magnetic order, either ferromagnetic~\cite{Lee2016} or antiferromagnetic. In the specific case of NiPS$_3$, bulk crystals display an unconventional zig-zag antiferromagnetic configuration~\cite{Joy1992,Wildes2015}, which persists in quasi-two-dimensional multilayer exfoliated samples~\cite{Kuo2016}.
Possible applications of NiPS$_3$ to optoelectronics
~\cite{Kim2018,Ho2021} and spintronics~\cite{Kim2019,Kang2020,Hwangbo2021,Wang22} have been proposed.

From a theoretical perspective, the magnetic behavior of NiPS$_3$ is commonly described using a generalized Heisenberg model with spin $S=1$ moments, associated with the Ni$^{2+}$ ions arranged on a honeycomb lattice~\cite{Olsen2021}. The low-energy excitation spectrum is largely captured by conventional spin-wave theory~\cite{Scheie2023}, and experiments have revealed that NiPS$_3$ also hosts significant coupling between magnetic order and optical excitations~\cite{Kang2020,Kim2018,Kim2023,Wang2021,Dirnberger2022,Jana2023,He2024,Allington2025,Wang25}.
 
Optical and resonant inelastic X-ray scattering (RIXS) experiments revealed the existence of two very narrow peaks near 1.5 eV at $k=0$ in NiPS$_3$, separated by a small energy gap of 22 meV.  The focus was put on the lowest energy one, called peak I, since it was the narrowest. It was interpreted as a magnetic exciton, somehow related to the antiferromagnetic zig-zag order, since it disappears above the N\'eel temperature in optical experiments. More precisely, its nature was assigned \cite{Kang2020} to that of a Zhang-Rice \cite{Zhang1988} exciton  formed between a hole in the $d$ shell of a Ni atom and another in the $p$ shell of the ligand anions, corresponding then to the $d^9 \underline{L}$ configuration. Since then, the nature of this excitation has been subject of debate. 
Moreover, a second excitation that lies only 22 meV higher in energy has received less attention, but it also disappears at the N\'eel temperature in optical experiments. It has been interpreted as a two-magnon sideband \cite{Kang2020} or an infrared-active phonon \cite{Schue26}
associated with peak I . The interest in these excitation peaks, as well as other structures below the conduction gap, has increased in the past few years \cite{Hwangbo2021,Ho2021,Wang2021,Belvin2021,Afanasiev2021,Dirnberger2022,KimKim2023,DiScala2023,Jana2023,Kim2023,Allington2025,Klaproth2023,He2024}.

Several calculations with interacting models in NiS$_6$ clusters indicate that the ground state is a triplet with mixed $d^8$ and $d^9\underline{L}$ character \cite{Kang2020,DiScala2023,Belvin2021,Klaproth2023,KimKim2023,He2024}, while the excited state associated with peak I has singlet character, with a larger contribution from the \(d^9\underline{L}\) configuration. Recent NMR experiments \cite{Kim26} confirm the presence of ligand holes in the singlet states, consistent with this picture and also with our DFT-based maximally localized Wannier functions, displayed in Fig. \ref{orbitales}, adopted in our approach.

In NiPS$_3$, the Zhang–Rice terminology has also been extended to the triplet ground state, commonly referred to as a Zhang–Rice triplet (ZRT) \cite{Kang2020}. The presence of the $d^9\underline{L}$ component also in the excited singlet is, in principle, consistent with a Zhang–Rice singlet (ZRS).  In the original ZR picture for cuprates, the singlet is formed by a $d$-hole and a ligand hole of the same symmetry, and is stabilized by their hybridization. In NiPS$_3$, both $e_g$ orbitals are involved — $d_{z^2}$ and $d_{x^2-y^2}$ if a cubic environment around Ni is assumed as a first approximation, which carry different spatial symmetries — along with their respective ligand counterparts. Besides, the triplet GS is instead stabilized primarily by Hund's exchange within the $d^8$ configuration. This difference in the microscopic origin of the binding has motivated other descriptions of the low-energy excitations where the Zhang-Rice picture is replaced by other alternatives \cite{Hamad2024, He2024,Wang2021,Hwangbo2021,Jana2023,Dirnberger2022}.

To describe the correlated excitations in NiPS$_3$, it is important to take into account the interactions within the Ni $3d$ shell. Our approach, described below, incorporates these interactions while also including the effects of covalency involving both S and P atoms. Previous cluster calculations for NiS$_6$ clusters \cite{Kang2020}, include the relevant Coulomb and exchange interactions within the Ni $3d$ shell, and thus also account for the role of Hund's exchange in determining the multiplet structure and excitation energies. Their description is restricted, however, to an isolated cluster (not including the P atoms, whose role  we have shown to be important \cite{Hamad2024}) and although they provided an important understanding of the intrinsic nature of the excitations, they naturally do not capture the propagation and dispersion of these excitations in the crystal. Our approach instead allows these correlated singlet states to evolve dynamically through the lattice.

Recent RIXS studies on related Ni$^{2+}$ charge-transfer 
insulators have provided additional insight into the microscopic 
nature of these excitations. In particular, 
Occhialini {\it et al.}~\cite{Occhialini2024} showed that the sharp excitonic features observed in Ni-based layered compounds can be interpreted as spin-singlet $d-d$ multiplet excitations, 
primarily stabilized by intra-atomic Hund's coupling. 
Within this framework, these authors propose that the role of 
charge-transfer configurations ($d^9\underline{L}$) is mainly 
to enable a finite dispersion via ligand-mediated hopping 
processes, rather than to define the exciton itself. 
Their results provide an alternative to Zhang-Rice-based interpretations and support a picture in which the excitations retain a predominantly local multiplet character.

The exciton dispersion provides direct insight not only into the microscopic mechanisms governing its propagation —such as virtual charge-transfer processes, ligand-mediated hopping pathways, or magnon-assisted motion— but also into the nature of the excitonic quasiparticle itself. In van der Waals transition-metal halides and related two-dimensional Mott and charge-transfer insulators, excitons often emerge as composite objects strongly entangled with orbital and spin degrees of freedom, rather than as purely local multiplet excitations. Consequently, the momentum dependence of the exciton energy constitutes an important test for competing theoretical models. 

In the particular case of NiPS$_3$, recent resonant inelastic X-ray scattering (RIXS) experiments~\cite{He2024} have revealed that the exciton dispersion closely follows that of the two-magnon excitations, despite the large separation in their absolute energy scales. This observation suggests that the microscopic processes governing exciton propagation are not purely electronic, but are instead intrinsically 
related to the antiferromagnetic background dynamics. In particular, it indicates that 
the motion of the exciton involves processes that are closely related to those 
underlying two-magnon excitations, pointing to a common origin in spin-exchange mechanisms.
  
In this work, using the previously developed theory for the singlet polaron excitations in NiPS$_3$, we calculate the dispersion of the quasiparticle excitations that naturally appear within the theory, and reconcile it with the experimentally measured dispersion by RIXS. We show that the experimental dispersion is well described by the model, and discuss how the calculated dispersion can be interpreted experimentally. Our theory is, so far, the only one that can calculate the dynamics of the excitons, and therefore it is important to discuss the results obtained from it in relation to experimental results.    

This paper is organized as follows. In the next section we review the singlet polaron theory (SPT) in NiPS$_3$, first presented in Ref. \cite{Hamad2024}, highlighting its main ingredients. 
In Sec.~\ref{dispersion} we present the calculated spectral functions and extract the 
dispersion of the singlet polarons, discussing their evolution across the Brillouin 
zone and comparing the results with RIXS experiments. Particular emphasis 
is placed on the relation between the exciton dispersion and the two-magnon continuum, 
showing that the former is largely controlled by the lower edge of the latter. 
Finally, in the last Section, we discuss the overall physics and point out some similar systems where this theory might apply.

\section{Singlet polaron theory in NiPS$_3$} 
\label{theory}

NiPS$_3$ crystallizes in a monoclinic structure with point group C$_{2h}$.
Each layer has a thickness of 3.18 Å and is composed of three atomic planes, while the interlayer distance is 3.73 Å. The Ni atomic plane is sandwiched between two S planes located above and below it. 

The magnetic structure within the Ni planes consists of ferromagnetically ordered zigzag chains running along the crystallographic $a$ direction, while these chains are coupled antiferromagnetically along the perpendicular $b$ direction (see Fig. 1 of Ref.~\cite{Wildes2015}).
As in most van der Waals materials, the interactions between atoms belonging to different layers (in the $c$ direction) can be considered negligible. Nevertheless, our density functional theory (DFT) calculations were performed using the full bulk structure \cite{helman}.

\begin{figure*} [tb]
\centering
\begin{subfigure}{0.4\textwidth}
\includegraphics[scale=0.8]{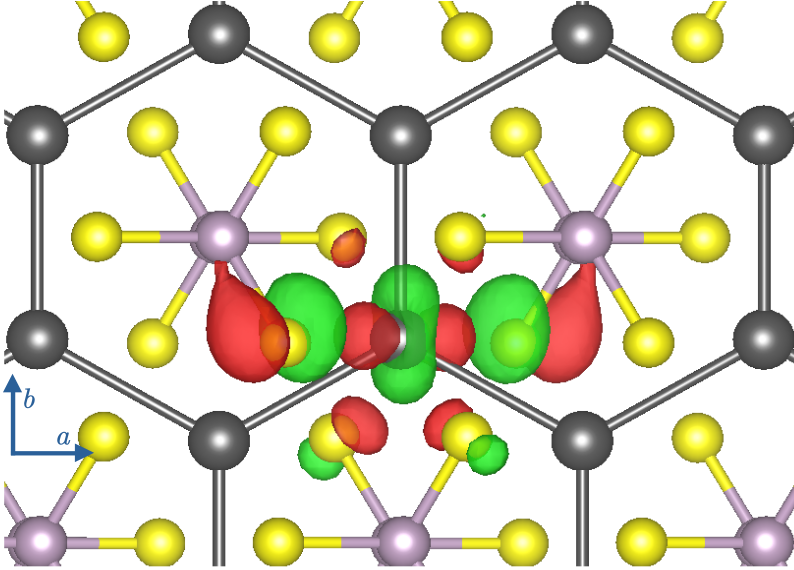}
\end{subfigure}
\quad
\begin{subfigure}{0.48\textwidth}
\includegraphics[scale=0.8]{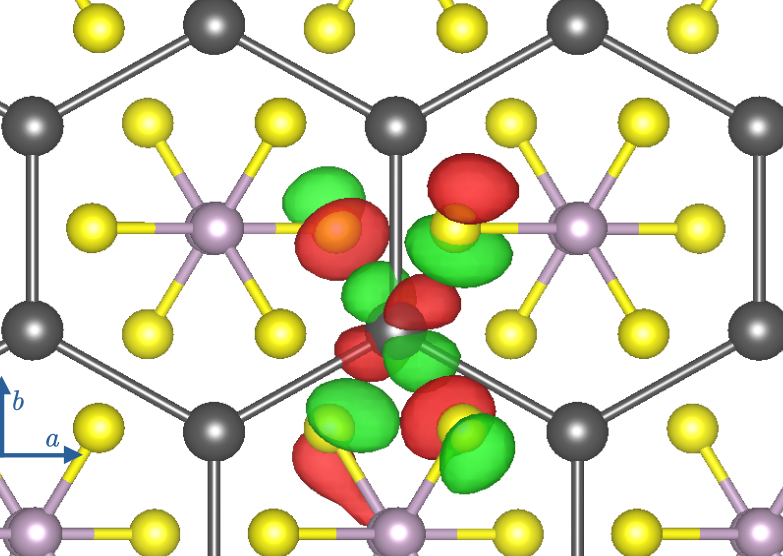}
\end{subfigure}
\caption{\label{orbitales} Top view of a single NiPS$_3$ layer. The hexagonal lattice is formed by Ni atoms (gray), while a P$_2$S$_6$ (yellow and violet, respectively) complex anion is located at the center of each hexagon. The left and right panels display the effective Wannier orbitals with even ($e$) and odd  ($o$) symmetry, respectively.
}
\end{figure*}

In order to construct a Hubbard Hamiltonian, a non-spin polarized DFT calculation was performed. Using maximally localized Wannier functions, and focusing on an energy window of 1.5 eV around the Fermi level,  
we obtain two Wannier functions centered at each Ni atoms, with an evident d-like character, but extending towards the nearest-neighbor S and P atoms. 
These orbitals are shown in Fig. \ref{orbitales}. They possess well-defined symmetry, and are labeled as $o$ and $e$, corresponding to ``odd'' and ``even'', respectively.
These molecular-like orbitals constitute the relevant degrees of freedom for magnetic excitations and fit well the bands in the energy window around the Fermi level. 

The obtained hopping parameters show that third-neighbor hopping is the dominant contribution due to the 'complex anion' constituted by S-P atoms.
It is important to note that the effective orbitals obtained in our construction naturally account for the long-range hopping processes that characterize this material. By contrast, more localized models usually retain only nearest-neighbor Ni–S hopping terms, thereby failing to properly capture the third-neighbor magnetic interactions mediated by the complex anion.

\subsection{Two-band Hubbard model on the honeycomb lattice}
 
A low-energy two-band Hubbard model was constructed, including standard on-site interactions for the two degenerate effective $d$ orbitals presented in the previous section~\cite{Aligia2004}. We stress once more that these effective d orbitals contain an important admixture of ligand ions. However, the interaction terms are dominated by the d contribution, retaining the same form and symmetry as for purely d orbitals. Therefore, the present approach incorporates the effects of covalency while providing a more transparent description of the underlying physics.

Then, $H_{\text{Hub}} = H_{\text{hop}} + H_{\text{int}}$~\cite{Hamad2024}, where
\begin{eqnarray}
H_{\text{hop}} &=&\sum\limits_{i\in A,\sigma }[t_{1}o_{i+\delta _{1}\sigma
}^{\dagger }o_{i\sigma }+t_{2}\sum\limits_{\delta _{j}\neq \delta
_{1}}e_{i+\delta _{j}\sigma }^{\dagger }e_{i\sigma }  \notag \\
&&-t_{3}\sum\limits_{\delta _{j}\neq \delta _{1}}(-1)^{j}(e_{i+\delta
_{j}\sigma }^{\dagger }o_{i\sigma }+o_{i+\delta _{j}\sigma }^{\dagger
}e_{i\sigma })+\text{H.c.}],  \notag \\
H_{\text{int}} &=&\sum\limits_{i}[U\sum\limits_{\alpha }n_{i\alpha \uparrow
}n_{i\alpha \downarrow }
+U^{\prime }\sum\limits_{\sigma }n_{io\sigma }n_{ie\bar{\sigma}}  \notag \\
&&+(U^{\prime }-J_H)\sum\limits_{\sigma }n_{io\sigma }n_{ie\sigma
}-J_H(o_{i\uparrow }^{\dagger }e_{i\downarrow }^{\dagger }e_{i\uparrow
}o_{i\downarrow }+\text{H.c.})  \notag \\
&&+J^{\prime }(o_{i\uparrow }^{\dagger }o_{i\downarrow }^{\dagger
}e_{i\downarrow }e_{i\uparrow }+\text{H.c.})],  \label{hhub}
\end{eqnarray}
 the first sum is over one of the two sublattices (A) of the honeycomb
lattice, $\alpha $ denotes the orbitals $o$ or $e$, $n_{i\alpha \sigma
}=\alpha _{i\sigma }^{\dagger }\alpha _{i\sigma }$ and $\bar{\sigma}=-\sigma 
$.
The eigenstates of the purely local Hamiltonian $H_{\text{int}}$ at the site $i$ are
\begin{eqnarray}
|ia\rangle  &=& \frac{1}{\sqrt{2}}\left( o_{i\uparrow}^{\dagger} o_{i\downarrow}^{\dagger} + e_{i\uparrow}^{\dagger} e_{i\downarrow}^{\dagger}
\right)|0\rangle, \quad E_{a} = U + J^{\prime},  \notag \\
|ib\rangle  &=& \frac{1}{\sqrt{2}}\left( o_{i\uparrow}^{\dagger} o_{i\downarrow}^{\dagger} - e_{i\uparrow}^{\dagger} e_{i\downarrow}^{\dagger}
\right)|0\rangle, \quad E_{b} = U - J^{\prime},  \notag \\
|ic\rangle  &=& \frac{1}{\sqrt{2}}\left( o_{i\uparrow}^{\dagger} e_{i\downarrow}^{\dagger} - o_{i\downarrow}^{\dagger} e_{i\uparrow}^{\dagger}
\right)|0\rangle, \quad E_{c} = U^{\prime} + J_H,  \notag \\
|it1\rangle &=& o_{i\uparrow}^{\dagger} e_{i\uparrow}^{\dagger}|0\rangle, \quad E_{t} = U^{\prime} - J_H,  \label{ee}
\end{eqnarray}
where only the triplet component with the highest projection $S_z = 1$ is written explicitly, and $J_H$ denote the Hund's coupling. For cubic symmetry, $J^{\prime} = J_H$ and $U^{\prime} = U - 2J_H$, so
the ground state is a triplet, in agreement with cluster calculations and experimental results. The two degenerate singlet excitations $b$ and $c$ lie $2J_H$ above the ground
state, while the singlet $a$ lies at $4J_H$. For the value $J_H = 0.75$~eV, which is well justified in Ref.~\cite{Hamad2024}, singlets $b$ and $c$ carry an excitation energy of $1.5$~eV, precisely matching
the excitonic resonances observed in optical and RIXS experiments. Singlet $a$, with excitation energy $4J_H \approx 3$~eV above the optical gap, is disregarded throughout. The degeneracy between $b$ and $c$ is lifted first by the departure from high cubic symmetry (implying small but nonzero $J^{\prime} - J_H$ and $U^{\prime} - U + 2J_H$), and further by the dynamics introduced through $H_{\text{hop}}$, as discussed below.

\subsection{Computation of singlet dynamics}
\label{sec:SCBA}

In this subsection, we summarize the many-body calculation for the singlet
dynamics (see Appendix~\ref{app:Heff} and Ref.~\cite{Hamad2024}
for further details). The dynamics of the excitations $b$ and $c$ is governed by
an effective $t$--$J$ Hamiltonian derived from $H_{\text{Hub}}$ using second-order
perturbation theory in the hopping processes [Eq.~(\ref{HtJ})].
As shown in Appendix ~\ref{app:Heff}, the effective Hamiltonian reads
\begin{equation}
H_{\rm eff} = H_{J} + H_t^{(0)} + H^{\rm int}_t,
\label{eq:hamil}
\end{equation}
where $H_J$ describes the magnetic background in terms of $S=1$ spin-wave (magnon) excitations, $H_t^{(0)}$ accounts for the free hopping of the singlets along the ferromagnetic chains of the zig-zag order, and $H^{\rm int}_t$ describes magnon-assisted hopping processes in which singlet motion disturbs the magnetic background and creates or destroys pairs of magnons~\cite{Hamad2006,Hamad2008}. While in the traditional SCBA only one magnon is generated by the movement of a hole through a spin-1/2 antiferromagnetic background, here two magnons are generated by the movement of the singlets because the background corresponds to spins 1. The spectral function of the singlet quasiparticles is computed using a generalized self-consistent Born approximation (SCBA), as detailed below. The explicit form of $H_{\rm eff}$, its derivation from $H_{\rm Hub}$, and a
sketch of the processes it contains are given in Appendix~\ref{app:Heff} and Fig.~\ref{fig:cartoon}.

\subsubsection{Magnon excitations}
The magnetic Hamiltonian $H_J$ contains anisotropic XXZ exchange interactions between first-, second-, and third-nearest-neighbor (NN) spins on the honeycomb lattice, together with single-ion  anisotropy contributions~\cite{Scheie2023}. We adopt the parameter values $J_1 = -2.7$~meV, $J_2 = -0.42$~meV, $J_3 = 13.9$~meV, $\Delta = 1$, $D_1 = -0.01$~meV, and $D_2 = 0.21$~meV from Ref.~\onlinecite{Scheie2023}, the most comprehensive study of the spin-wave spectrum of NiPS$_3$ to date. Since third-NN antiferromagnetic correlations dominate, the classical ground state is the zig-zag order with magnetic wavevector $\bm{Q} = (0, 2\pi/3a)$.

We apply linear spin-wave theory (LSWT) to $H_J$ using the Holstein-Primakov representation~\cite{Toth2015}. After a local spin rotation that maps the zig-zag order onto a ferromagnetic one, we obtain a quadratic bosonic Hamiltonian in momentum space with two sublattices $A$ and $B$ per unit cell, written compactly as
\begin{equation}
H_J = \frac{1}{2}\sum_{\bm k}\bm{a}_{\bm k}^\dagger \cdot \mathcal{M}_{\bm k} \cdot \bm{a}_{\bm k},
\end{equation}
where the bosonic spinor is $\bm{a}_{\bm k}^\dagger = (a^\dagger_{\bm k A},\, a^\dagger_{\bm k B},\,a_{-\bm k A},\, a_{-\bm k B})$ and $\mathcal{M}_{\bm k}$ is a $4\times 4$ dynamical matrix containing both
normal and anomalous intra- and inter-sublattice terms (explicit form in
Appendix~\ref{app:Heff}). The two magnon branches $\omega_{\bm k\eta}$ ($\eta = 1, 2$ denote the two branches due to the two-sublattice structure) \footnote{If the local rotation is not performed, the magnetic unit cell contains four atoms, resulting in four branches within the reduced Brillouin zone, as in the treatment of Wildes et al.~\cite{Wildes2022}.} and the corresponding Bogoliubov operators $\alpha_{\bm k\eta}$ are obtained by paraunitary diagonalization of $\mathcal{M}_{\bm k}$ following Ref.~\onlinecite{COLPA1978}, yielding a transformation matrix $\mathcal{F}_{\bm k}$ that satisfies the paraunitary condition
$\mathcal{F}_{\bm k}^\dagger \cdot \mathcal{S} \cdot \mathcal{F}_{\bm k} = \mathcal{S}$, with $\mathcal{S} = \mathrm{diag}(1,1,-1,-1)$.

\subsubsection{Singlet hopping and magnon-assisted vertices}
We define singlet annihilation operators $h_{\bm k \xi \alpha}$ 
($\xi = A, B$ labels the two sublattices, and $\alpha = b, c$ 
labels the two singlet flavors corresponding to the destruction 
of $b$ and $c$ singlets). These operators enter the bare hopping term $H_t^{(0)}$ through a $4\times 4$ matrix $\mathcal{M}^h_{\bm k}$, which describes free propagation along ferromagnetic bonds, 
i.e., first-nearest-neighbor hopping within the zigzag chains. Third-nearest-neighbor
bonds, which carry the dominant hoppings, all connect antiparallel spins and
therefore contribute only to the magnon-assisted term $H_t^{\rm int}$ (see
Appendix~\ref{app:Heff} and Table~\ref{tab:hop}).
Note that the hopping matrix is not diagonal in the flavor indices $\alpha$,
so the singlet can change character as it propagates.

The magnon-assisted term $H^{\rm int}_t$ couples singlet hopping
to the creation or annihilation of magnon pairs. Its vertex
$M^{\xi\xi',\eta\eta'}_{\alpha\beta}(\bm k,\bm q,\bm q')$, for a singlet of flavor
$\beta$ and momentum $\bm k$ that leaves sublattice $\xi$ and reaches sublattice
$\xi'$ as flavor $\alpha$ while the magnons $(\bm q,\eta)$ and $(\bm q',\eta')$
are emitted, is derived in Appendix~\ref{app:Heff}, Eq.~(\ref{vertexA}).

By 'magnon-assisted propagation' we refer to the coherent process by which the singlet exciton moves through the antiferromagnetic background by creating and reabsorbing magnons. Although bare hopping along the ferromagnetic bonds is also allowed, in closely analogous systems \cite{Hamad2006, Hamad2007} we have shown that the quasiparticle is primarily formed by the magnon-assisted processes, while the bare hopping contributes to spectral weight at higher energies.

\subsubsection{Self-consistent Born approximation}
The singlet spectral function is computed using the self-consistent Born approximation (SCBA)~\cite{Kane1989}, which resums the dominant class of diagrams corresponding to a singlet that propagates while emitting and reabsorbing a pair of magnons  [see Fig.~\ref{fig:cartoon}(d)]. The matrix Dyson equation for the $4\times 4$ singlet Green function is
\begin{equation}
\bm{G}^{-1}(\bm k,\omega) = \bm{G}_0^{-1}(\bm k,\omega) - \bm{\Sigma}(\bm k,\omega),
\end{equation}
where the bare inverse propagator is $(G_0^{-1})^{\xi\kappa}_{\alpha\beta} =
(\omega + i0^+ - \varepsilon_\alpha)\delta_{\xi\kappa}\delta_{\alpha\beta} -
(\mathcal{M}^h_{\bm k})^{\xi\kappa}_{\alpha\beta}$, with
$\varepsilon_\alpha=E_\alpha-E_t$,
and the self-energy is given by 
\begin{widetext}
\begin{equation}
\Sigma_{\alpha\beta}^{\xi\kappa}(\bm k,\omega) = \sum_{\bm q,\bm q'}\sum_{\alpha'\beta'}\sum_{\xi'\kappa'}\sum_{\eta\eta'}
M^{\kappa'\kappa,\eta\eta'}_{\beta'\beta}(\bm k,\bm q,\bm q')\, M^{*\xi\xi', \eta\eta'}_{\alpha'\alpha}(\bm k,\bm q,\bm q')\, G_{\alpha'\beta'}^{\xi'\kappa'}\!\left(\bm k{-}\bm q{-}\bm q',\,
\omega - \omega_{\bm q\eta} - \omega_{\bm q'\eta'}\right).
\label{eq:selfenergy}
\end{equation}
\end{widetext}
This equation is solved self-consistently by iteration until convergence. The spectral density for each singlet flavor $\alpha$ on sublattice $\xi$ is then obtained from the imaginary part of the corresponding diagonal element of the Green function:
\begin{equation}
A_{\alpha\xi}(\bm k,\omega) = -\frac{1}{\pi}\,\mathrm{Im}\,G_{\alpha\alpha}^{\xi\xi}(\bm k,\omega).
\end{equation}

\begin{figure}[h]
\includegraphics[width=0.9\linewidth]{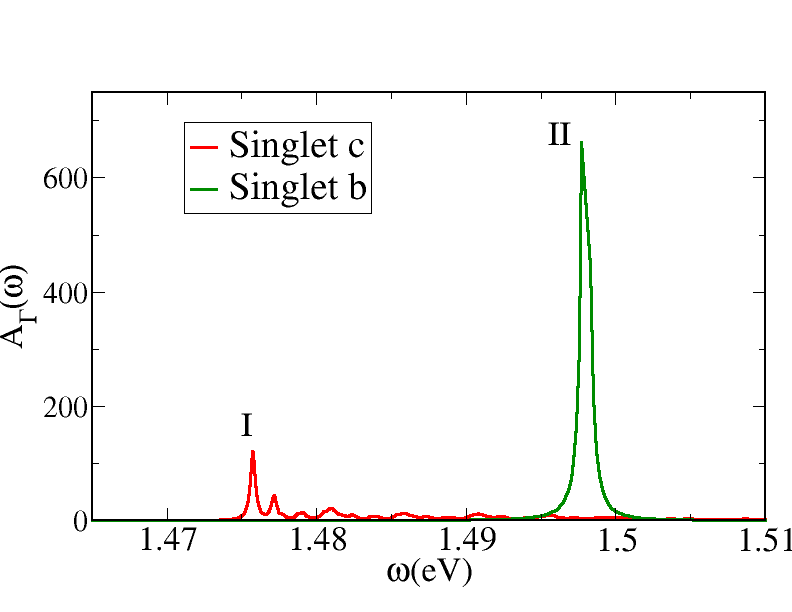}
\caption{(Color online) 
Spectral function at $\bm{k} = (0,0)$, reproduced from Ref.~\cite{Hamad2024}. 
The red and green curves correspond to singlets $c$ and $b$, respectively; the full spectral function is 
their superposition.}
\label{spectrum_gamma}
\end{figure}

\begin{figure*}[t]
    \centering
    \begin{subfigure}{0.48\textwidth}
        \includegraphics[width=\linewidth]{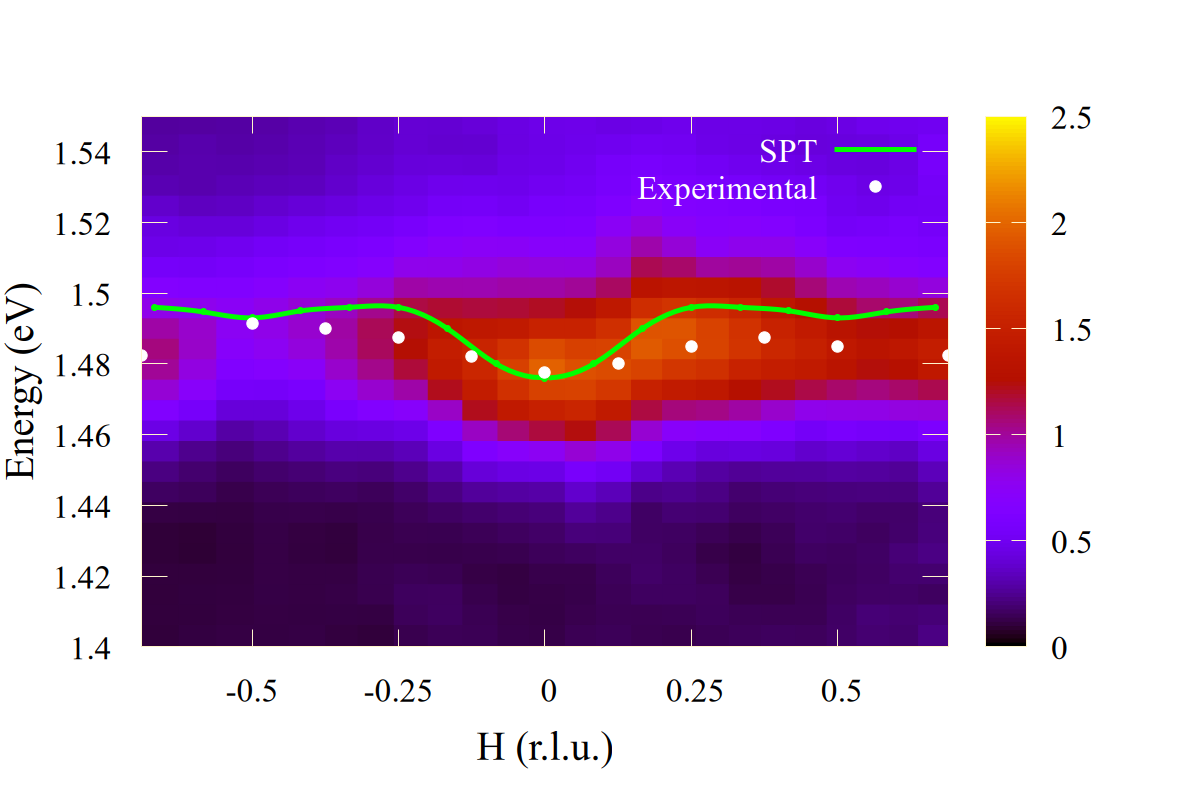}
            \label{mapacolor_H}
    \end{subfigure}\hfill
    \begin{subfigure}{0.48\textwidth}
        \includegraphics[width=\linewidth]{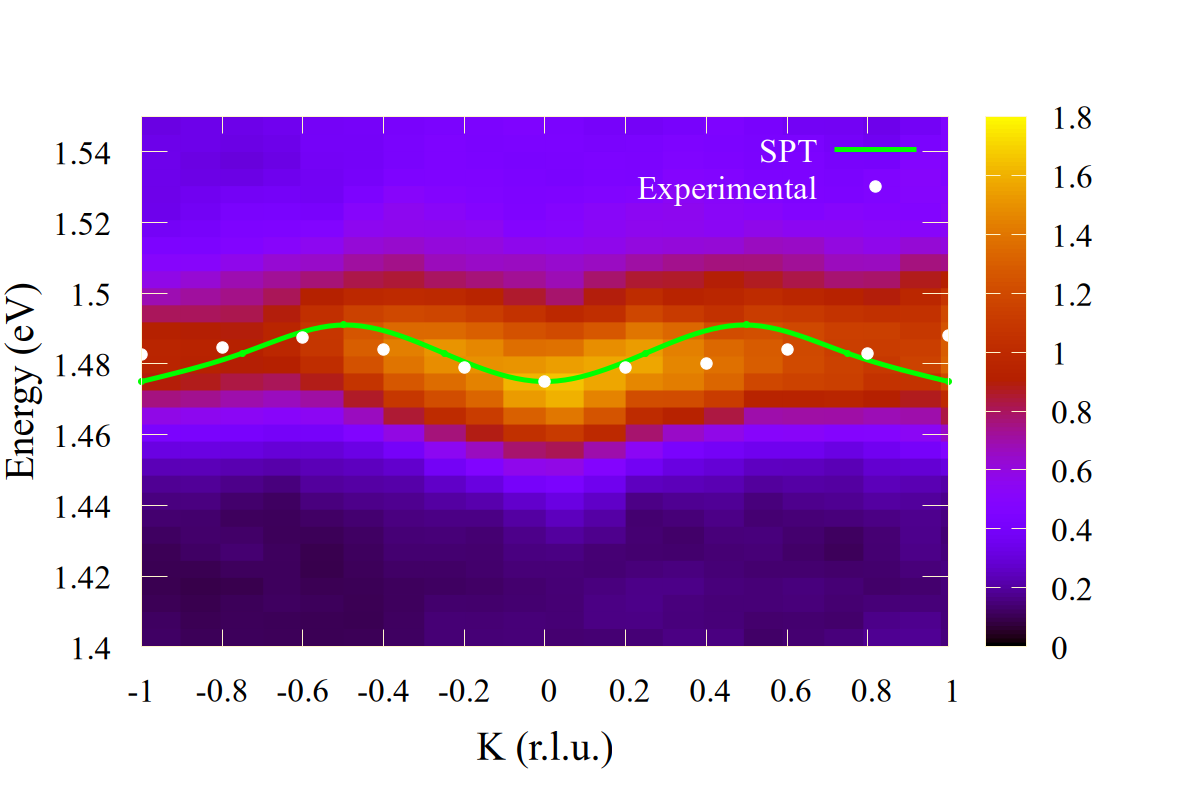}
        \label{mapacolor_K}
    \end{subfigure}\hfill
    \caption{(Color online) Comparison with experiment. Experimental RIXS intensity maps  from Ref.~\cite{He2024} along the $H$ (left) and $K$ (right) high-symmetry directions of the Brillouin zone.  The color scale encodes the measured scattering intensity (in arbitrary units). White dots, taken from Ref.~\cite{He2024}, indicate the peak positions of the excitonic resonance fitted from the data and the green curves correspond to the singlet polaron theory (SPT) dispersion relation (see text for details). }
    \label{disp_compared}
\end{figure*}

\subsection{Comparison with optical experiments}

The spectral function at the $\Gamma$ point, directly comparable to optical experiments, is reproduced in Fig.~\ref{spectrum_gamma} from Ref.~\cite{Hamad2024}. The two peaks, corresponding to singlets $c$ and $b$ respectively, arise from the interplay between the bare singlet energies and their renormalization due to their coupling to the magnon bath. This picture provides a natural interpretation of the optical resonances in
NiPS$_3$ as \textit{singlet polarons}: quasiparticles whose main character is ionic (their energy scale is set by $J_H$), but whose spectral manifestation is shaped by their dynamics and coupling to magnetic excitations. This is a conceptually distinct alternative to the Zhang-Rice exciton picture, which attributes the lowest optical resonance to a single excitation delocalized between the Ni $d$ orbital and the ligands.

It is worth discussing the symmetry assignment of the observed singlet excitations. Previous cluster calculations have assigned peak I to a \(^{1}A_{1g}\) singlet state, while peak II has been interpreted as a sideband of this excitation, possibly involving double-magnon or phonon processes. Moreover, the spectral feature at around 1 eV has been associated with a singlet with \(^{1}E_g\) symmetry. In the present approach, however, the \(^{1}E_g\) sector is spanned by the two states \(|i_b\rangle\) and \(|i_c\rangle\), which are degenerate in the ideal \(O_h\) limit. In the crystal, this degeneracy can be lifted by deviations from the ideal octahedral symmetry and, importantly, by the dynamics of the singlet excitations, leading to two closely spaced excitonic states, as discussed in the previous work \cite{Hamad2024}. We therefore associate these two states with peaks I and II. In contrast, the \(^{1}A_{1g}\) state \(|i_a\rangle\) lies at higher energy in our model and therefore cannot be identified with the feature around \(1\,\mathrm{eV}\). We also note that the latter feature has been assigned in previous studies to \(d\)-\(d\) excitations \cite{Klaproth2023} , rather than to a singlet–triplet excitation. 

We note that this interpretation of peaks I and II associated with singlets $|b\rangle$ and $|c\rangle$ is also consistent with the more recent observation of level repulsion between peaks I and II under an applied magnetic field [Ref. \cite{Huang2026}], where the two features are interpreted as two closely related excitonic states. Furthermore, the singlet polaron picture is also consistent with the results of Ref.~\cite{Allington2025}, that demonstrated that the creation of these excitons in NiPS$_3$ naturally induces long-wavelength magnons, in direct agreement with the two-magnon processes that are central to our theoretical description.

A more stringent test of the theory is provided by the momentum-resolved dispersion
of these excitations, accessible through RIXS. The comparison between our SCBA results and RIXS measurements is the central result of the present work, and is presented in Sec.~\ref{dispersion}.

\section{Singlet polaron dispersion and comparison with RIXS}
\label{dispersion}

Because of the inherent properties of photons, optical experiments using visible light are restricted to probing excitations near the center of the Brillouin zone (the $\Gamma$ point), corresponding to wave vector $\bf k=0$, though they offer very high energy resolution. In contrast, resonant inelastic X-ray scattering (RIXS) experiments provide lower energy resolution but enable measurements of excitation
dispersions across the entire Brillouin zone. Due to this intrinsically lower resolution, even the most accurate RIXS experiments can only see a broad spectral weight around $1.5$~eV, but with the proper analysis clear hints of a two-peak structure can be inferred from the data~\cite{Kang2020}.

The experimental RIXS intensity maps along the $H$ and $K$ directions, reproduced in Fig.~\ref{disp_compared} from Ref.~\cite{He2024}, show this broad feature centered near $1.5$~eV. 
The peak positions extracted from the data reveal a clear momentum dependence, with a total bandwidth of approximately $15$~meV ($12$~meV) along the $H$ ($K$) direction.

Fig.~\ref{disp_compared} shows the dispersion of the singlet polarons of our model, compared with the peak positions extracted from the RIXS data. 
To understand this comparison, we first note that within our theory there are two singlet polarons; accordingly, the spectral function $A(\mathbf{k},\omega)$ consists in principle of 
two quasiparticle (QP) peaks, each associated with one of them. In addition to these QP peaks, $A(\mathbf{k},\omega)$ contains satellite peaks and an incoherent multi-magnon continuum at 
higher energies.

\subsection{Momentum evolution and structure of the spectral function}

\begin{figure*}[t]
    \begin{subfigure}{0.5\textwidth}
        \includegraphics[scale=0.7]{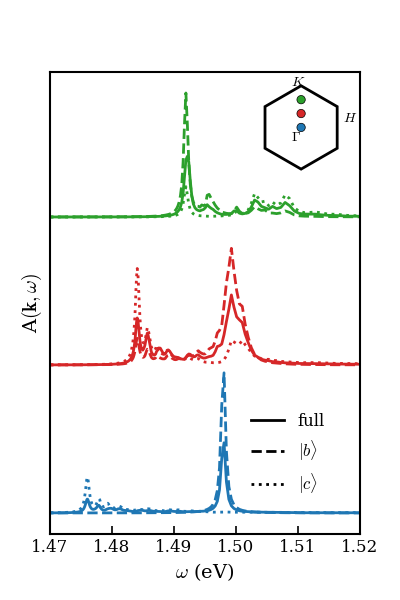}
        \caption{\label{Spectra_K}}
    \end{subfigure} \hfill
    \begin{subfigure}{0.48\textwidth}
        \includegraphics[scale=0.7]{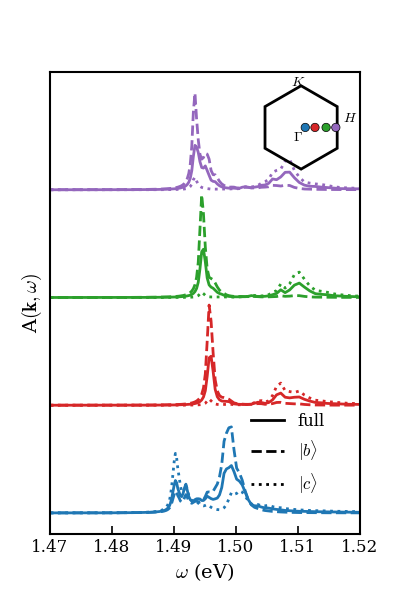}
        \caption{\label{Spectra_H}}
    \end{subfigure}
    \caption{(a)Evolution of the spectra along the $K$ direction. Apart from the $\Gamma$ point, there are two different spectral functions for our cluster size (N=192). The QP peak and also other peaks of the partial spectral functions of each singlet polaron merge.
    (b) Evolution of the spectra along the $H$ direction. Dashed/dotted line corresponds to singlet $b$ ($c$). The QP peak and also other peaks of the partial spectral functions of each singlet polaron merge.}
    
\end{figure*}

While at $k=(0,0)$ the two QP peaks are clearly separated (the lowest energy one being 
the one associated with singlet $c$), as long as the spectral function is evaluated 
away from the $\Gamma$ point, both in the $K$ and $H$ directions, the spectral functions obey a general trend with  these two QP peaks merging and forming a unique low energy peak. 
Moreover, other peaks or structures at higher energy, corresponding to the incoherent part 
of the spectral function, also show coincidence in energy, i.e. the peaks observed for 
the spectral function can no longer be attributed to singlet $c$ or $b$ alone, unlike what happens 
for the $\Gamma$ point. 
However, for some momenta, the relative intensity of each partial spectral function can be very different, especially regarding the QP peak. 
This can be seen for each direction along the Brillouin zone, in Figs.~\ref{Spectra_K} and \ref{Spectra_H}. 
At momenta closer to the $\Gamma$ point, the QP peak has mainly a singlet $c$ character, but with some small component of singlet $b$; however, as the spectral function is evaluated at 
momenta further away from the $\Gamma$ point, the QP peak gradually gains contributions from singlet $b$.

The structure of the spectral functions obtained within SCBA 
(see Figs.~\ref{Spectra_K} and \ref{Spectra_H}) offer further insight into the nature of the exciton quasiparticle. In addition to the coherent quasiparticle peak, the spectra display a broad incoherent background arising from multi-magnon excitations. This continuum reflects the fact that the exciton propagates by emitting and 
reabsorbing pairs of magnons, leading to a strong dressing of the singlet excitations. As a consequence, the position of the lowest-energy peak cannot be obtained from a model where the singlets move freely, but rather it is the result of a non-trivial self-energy renormalization governed by the magnetic excitation spectrum.

The dispersion of the lowest energy peak in our theory is shown in Fig.~\ref{disp_compared}. 
All the spectral functions were calculated for a lattice of $N=192$ unit cells. 
It can be seen that our theory reproduces well the dispersion obtained 
from the RIXS experiments. The bandwidth of the RIXS dispersion along the $H$ ($K$) direction is of about 15 meV (12 meV), being the one from our theory slightly larger, 
of about 17 meV (16 meV), respectively. The calculated dispersions are even with respect to 
the $\Gamma$ point, a trend that is also consistent with the experimental dispersions, 
within the limits of experimental uncertainty.

It could be argued that since RIXS experiments are not as precise as optical ones, instead of the several peaks in our spectral functions, a broad peak encompassing the entire spectrum would be observed instead. In any case, probably the most intense part of the spectrum would be marked as the energy of the exciton, since most works so far have considered a single exciton in this material. However it should be noted that the relative intensities of the spectral function $A(k,\omega)$ and of the spectra in optical or RIXS experiments are not necessarily the same, since they depend on matrix elements.  It is important to emphasize that the intrinsic experimental resolution of RIXS experiments hinder a detailed comparison with our results. While the theory predicts quasiparticle peaks and incoherent contributions, the finite experimental resolution leads to a broadening and partial merging of these features.

Therefore, the experimentally extracted dispersion should be understood as corresponding to the maximum of the spectral weight, rather than to a well-defined quasiparticle pole. In this context, the good agreement between the calculated dispersion of the lowest-energy peak and the experimental data indicates that the dominant spectral weight is indeed controlled by the same mechanisms captured within the SCBA.
This also explains why higher-energy structures present in the theoretical spectral function are not clearly resolved experimentally, but instead contribute to an overall broadening and asymmetry of the observed peak.
In any case, for example, if the dispersion was obtained from the position of the spectral center of mass rather than from the QP peak, it would have a similar shape but shifted upward in energy by approximately 2--3 meV.

It is worth noting that our model relies on several approximations. It is derived from a two-band Hubbard model for two orbitals, employs second-order perturbation theory to obtain an effective Hamiltonian, and is solved using a generalized SCBA. Although SCBA is among the most reliable methods for computing spectral functions of quasiparticles coupled to magnetic excitations, it is not exact. Nevertheless, it is important to emphasize that 
the only free parameters of the final effective model [Eq.~(\ref{eq:hamil}) and
Appendix~\ref{app:Heff}], the bare singlet energies $\varepsilon_b$ and
$\varepsilon_c$, have been kept fixed throughout the calculation of the spectral functions presented here, and we have assumed an on-site repulsion $U=4$ eV. We have computed the dispersion for larger values of $U$, finding that the bandwidth slightly decreases for increasing $U$, but the overall shape of the dispersion curve remains unchanged.  Furthermore, some discrepancies may arise from the intrinsically lower energy resolution of RIXS experiments compared to optical measurements, so that RIXS dispersion may not directly reflect the spectral features of the calculated spectral function.

We now comment on the temperature dependence of these excitations. In NiPS$_3$, RIXS experiments \cite{He2024}  show that the excitonic features become weaker and more diffuse above $T_N$ while 
no dispersion is detectable. In optical measurements of Ref. \cite{Kang2020}, the spectral weight changes more strongly through $T_N$, which He et al. \cite{He2024} attribute to the role of exciton-spin interactions 
in enabling the optical cross-section. This behavior is naturally explained within our picture: the local singlet energy ($\sim 2J_H\approx 1.5$~eV) far exceeds the thermal energy at $T_N$, so the excitation 
survives locally, while the magnon-assisted propagation — which requires long-range magnetic order — is lost. On the other hand, a finite exciton dispersion has also been reported in related 
Ni-based charge-transfer insulators~\cite{Occhialini2024}. In those systems, the dispersion has been attributed to ligand-mediated hopping processes associated with $d^9 \underline{L}$ 
configurations, and was found to persist even above the magnetic ordering temperature. These differences suggest that different microscopic mechanisms may govern exciton propagation in different materials. 
In particular, while in NiPS$_3$ the dynamics appears to be dominated by magnon-assisted processes, in more covalent systems ligand-mediated hopping may provide the leading contribution to the exciton dispersion.

We would also like to provide a plausible explanation for why the excitations observed around $1$ eV show essentially no dispersion. As mentioned above, these excitations are most likely internal $d-d$ excitations. Their energies, being well below $2J_H$, suggest that they correspond to spin-conserving excitations in which an electron is promoted between the $e_g$ and $t_{2g}$ manifolds. The resulting excited state therefore contains both an $e_g$ and a $t_{2g}$ hole.
For this excitation to propagate from site $A$ to site $B$, the $t_{2g}$ hole at $A$ must move to $B$, while an $e_g$ hole at $B$ must move to $A$. These processes arise only at second order in the $d-p$ hopping through the ligand orbitals. Furthermore, hopping involving the $t_{2g}$ orbitals is governed by the $(pd_\pi)$ matrix elements, which are substantially smaller than the $(pd_\sigma)$ matrix elements relevant for the $e_g$ orbitals.  We therefore expect the effective hopping, and hence the dispersion, of these $d-d$ excitations to be much smaller than that of the singlet excitations considered in our work. This provides a plausible explanation for why the corresponding RIXS features appear essentially dispersionless.

\subsection{Interplay between singlet polaron dispersion and the two-magnon continuum}
\label{sec:twomagnon_exciton}

\begin{figure}[htp]
\includegraphics[width=1.0\linewidth]{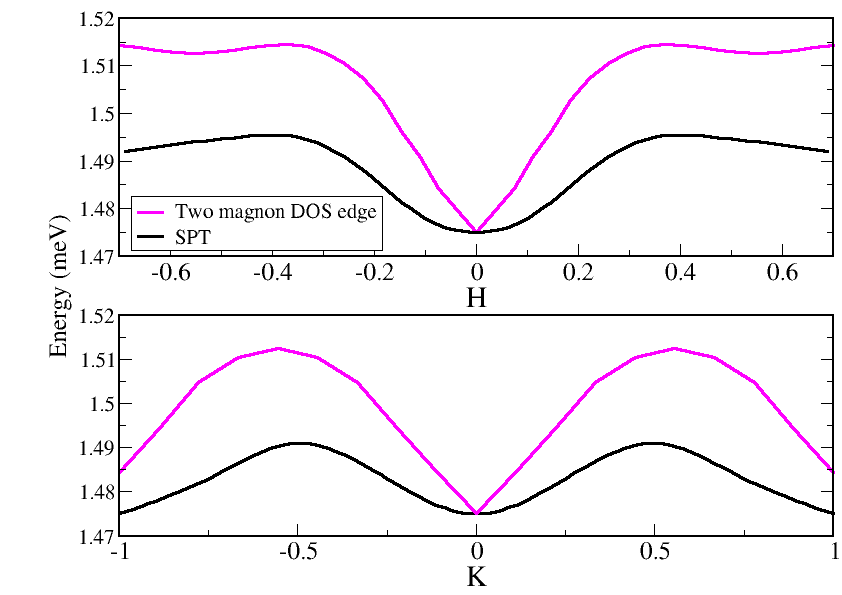}
\caption{(Color online) Comparison between the singlet polaron dispersion and the threshold for  the two-magnon continuum within the SCBA along the $H$ (top panel) and $K$ directions (bottom panel). The latter has been shifted in energy so as minima of both curves coincide.}
    \label{fig:twomagnon}
\end{figure}

The recent RIXS experiments have revealed a close relation between the exciton  
dynamics in NiPS$_3$ and magnetic excitations~\cite{He2024}. In particular, it was shown that the dispersion of the exciton closely follows that of the two-magnon excitation, despite the large separation in their absolute energy scales~\cite{He2024}. 
This observation strongly suggests that the mechanism governing exciton propagation is intrinsically related to magnetic exchange interactions in the antiferromagnetic background.

Microscopically, the motion of an exciton in a correlated magnetic background necessarily disturbs the magnetic configuration, generating strings of misaligned spins. Coherent propagation then requires higher-order processes that effectively cure these defects 
via multiple spin exchanges. As discussed in Ref.~\cite{He2024}, both the exciton and the two-magnon excitation propagate through sequences of spin-exchange processes of comparable structure, which naturally explains the similarity in their dispersions.

Within our singlet polaron framework, this connection arises in a more direct and quantitative way. The exciton quasiparticle is described as a singlet propagating in a $S=1$ antiferromagnetic background, strongly coupled to magnon excitations. In the SCBA treatment, its dynamics is governed by the emission and reabsorption of pairs of magnons, leading to a self-energy that is directly controlled by the two-magnon continuum. As a consequence, the dispersion of the quasiparticle is not only determined by local interaction scales (such as Hund's coupling), but also by the structure of the magnetic excitation spectrum.

A more detailed analysis reveals that the singlet polaron dispersion obtained within SCBA is closely related to the lower edge of the two-magnon continuum. This is explicitly shown in Fig.~\ref{fig:twomagnon}, where we compare the singlet polaron SCBA dispersion with the RIXS data and the estimated two-magnon threshold. The agreement is good, particularly near the zone center and along both high-symmetry directions. Notice the energy scale in the graph: while the bandwidths may differ, in absolute terms, measured in eV, they are of the same order of magnitude and the differences very small.

This behavior can be understood from the structure of the self-energy (see Eq.~\ref{eq:selfenergy}),
\begin{equation}
\Sigma(\mathbf{k},\omega) \sim \sum_{\mathbf{q},\mathbf{q}'} 
|M(\mathbf{k},\mathbf{q},\mathbf{q}')|^2 \,
G(\mathbf{k}-\mathbf{q}-\mathbf{q}',\, \omega - \omega_{\mathbf{q}} - \omega_{\mathbf{q}'}) ,
\nonumber
\end{equation}
where $\omega_{\mathbf{q}}$ is the magnon dispersion. Since the self-energy involves a convolution over pairs of magnons with total momentum fixed by the quasiparticle, the dominant contributions arise from states with minimal excitation energy, set by the minimum of $\omega_{\mathbf{q}}+\omega_{\mathbf{P}-\mathbf{q}}$, where
$\mathbf{P}=\mathbf{q}+\mathbf{q}'$ is the total momentum of the pair. This quantity defines the lower edge of the two-magnon continuum, and therefore imprints its momentum dependence onto the renormalized quasiparticle pole. In this sense, the exciton dispersion becomes effectively locked to the two-magnon threshold, rather than to the full spectral weight of the continuum.
This provides a natural interpretation of the experimental findings of Ref.~\cite{He2024}: the exciton peak observed in RIXS can be understood as a relatively sharp feature sitting at the onset of a broader magnetic continuum.

This observation also allows us to critically contrast different theoretical pictures. The Hund exciton mechanism proposed by He {\it et al.}~\cite{He2024}, shares many of the basic ingredients of our singlet polaron theory. In that case, the exciton is primarily viewed as a local atomic excitation whose propagation is enabled by spin-exchange processes. While this picture correctly captures the magnetic nature of the propagation, it does not offer a framework to make quantitative predictions that can be directly compared to experiments. On the other hand, our theory can explicitly account for the strong dynamical renormalization induced by the coupling to the continuum of magnetic excitations. 

Within our singlet polaron framework the coupling to two-magnon states is built in at the microscopic level and treated non-perturbatively within SCBA. As a result, the quasiparticle is not a simple propagating excitation dressed by small corrections, but rather a strongly renormalized object whose very existence and dispersion emerge from its interaction with the magnetic background. From this perspective, the agreement between the SCBA dispersion and the two-magnon edge is not accidental, but a direct manifestation of the underlying many-body dressing.

\section{Conclusions and perspectives}

The nature of the excitations observed in NiPS$_3$ by both optical and RIXS experiments continues to be scrutinized, both by experimental and theoretical means. In this work, we have shown that within the singlet polaron framework these excitations can be 
consistently understood as quasiparticles arising from two singlet states dispersing through the 
lattice and strongly coupled to magnetic excitations. So far, this is the only microscopic approach that 
simultaneously accounts for the internal structure of the exciton, its dynamics across the Brillouin zone, and its coupling to the magnetic background on equal footing.

Within the SCBA treatment, this naturally leads to a dispersion that closely follows the lower edge of the two-magnon continuum, providing a direct microscopic basis for the behavior observed in RIXS experiments~\cite{He2024}. In this sense, our results go beyond a qualitative association between excitons and magnetic excitations, showing instead that the exciton dispersion is quantitatively controlled by the two-magnon continuum itself.

A key aspect of our approach is that it relies on a single set of parameters, obtained from first-principles calculations~\cite{Hamad2024} and from careful analysis of the magnetic sector~\cite{Scheie2023}. 
Without further adjustment, the theory provides a consistent description of both the optical excitations~\cite{Kang2020} and the dispersion measured by RIXS~\cite{He2024}. 
This level of agreement across different experimental probes strongly supports the validity of the underlying microscopic picture.

Our results capture, both qualitatively and with good quantitative agreement, the dispersion observed experimentally, reproducing the position of the minima and the overall shape along both $H$ and $K$ directions. In particular, our theory reproduces the flatness of the dispersion measured experimentally.  The resulting dispersion is also qualitatively consistent with a simplified tight-binding picture with an effective third-nearest-neighbor hopping \cite{He2024}. However, the actual dynamics is much more complex.  

The intensities observed in absorption and photoluminescence experiments depend on matrix elements, which have not been addressed here. In particular, local $d-d$ excitations are dipole-forbidden in centrosymmetric environments, and triplet-singlet transitions are also suppressed, so that finite spectral weight requires spin-orbit coupling and hybridization between transition-metal d electrons and ligand orbitals. Instead due to the presence of intermediate states, RIXS allows  d-d transitions. Spin-flip processes are also allowed, due to the strong spin-orbit coupling of the intermediate states. There is a tacit assumption that these mechanisms are at play, as discussed recently~\cite{Matsu26}.

More generally, our results indicate that in correlated magnetic insulators exciton dynamics cannot be 
disentangled from the underlying spin degrees of freedom, opening the possibility of using 
excitons as probes of magnetic correlations in van der Waals antiferromagnets.

Finally, recent RIXS experiments on triangular-lattice nickel 
dihalides NiX$_2$ (X = Cl, Br, I) have reported excitations 
with similar energy scales and finite dispersion~\cite{Occhialini2024}. In those systems, the excitons have been identified as local spin-singlet $dd$ multiplet excitations, whose dispersion is primarily attributed to ligand-mediated hopping processes associated with $d^9\underline{L}$ configurations, and which persist even in the absence of long-range magnetic order. This comparison suggests a more general picture in which excitons in Ni-based correlated materials originate from local multiplet excitations, while their propagation may be governed by different microscopic mechanisms depending on the material. In NiPS$_3$, our results indicate that the dominant mechanism is magnon-assisted propagation, leading to a dispersion controlled by the two-magnon continuum. In contrast, in more covalent systems, ligand-mediated processes may play a leading role.  Clarifying the interplay between these mechanisms, and identifying the regimes in which each one dominates, constitutes an interesting direction for future work.

\paragraph*{Acknowledgements.}

We thank Mark Dean for illuminating discussions. LOM is supported by CONICET under grant no. 3220 (PIP2021). IJH is supported by CONICET under grant no. 0883 (PIP2021) and by Universidad Nacional de Rosario under grant no. 80020220700200UR. CSH is supported by PICT-2021-00325 of the Agencia I+D+i, Argentina. AEF is supported by the U.S. Department of Energy, Office of Science, Basic Energy Sciences under award No. DE-SC0022216. This work used computational resources from UNC Supercómputo (CCAD) – Universidad Nacional de Córdoba (https://supercomputo.unc.edu.ar), which are part of SNCAD, República Argentina.

\appendix

\section{Effective Hamiltonian for the singlet-polaron dynamics}
\label{app:Heff}

\begin{figure*}[tb]
\centering
\includegraphics[width=\textwidth]{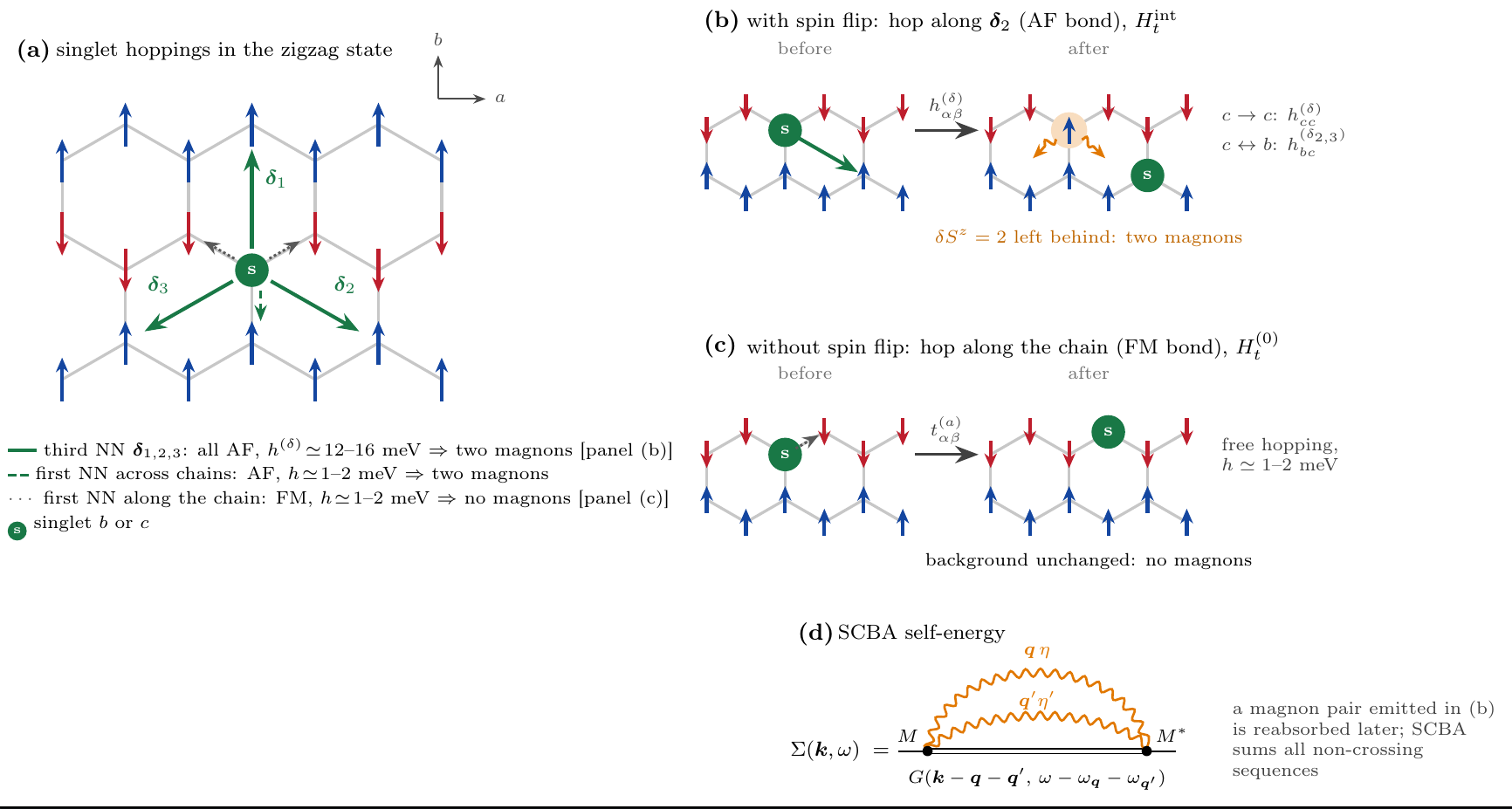}
\caption{\label{fig:cartoon}Processes included in the effective Hamiltonian. (a) NiPS$_3$ layer in the zigzag state.
From a singlet (green), the three third-NN bonds $\bm\delta$'s, which carry the 
dominant singlet hoppings, all connect antiparallel spins. 
The smaller first-NN hoppings connect parallel spins along the FM chain (dotted) and antiparallel spins across chains (dashed). 
(b) Hop with spin flip shown for the third-NN bond $\bm\delta_2$ (before and after): 
singlet and triplet interchange and the site left behind (shaded) 
ends up with its spin reversed with respect to the order, a deviation $\delta S^z=2$, i.e., two magnons. 
 The reverse process, in which the singlet enters a site that carries two
deviations due to zero-point fluctuations, removes them. (c) Hop without spin flip: the triplet arrives aligned with the order, the background is unchanged, and no magnons are emitted. 
(d) SCBA self-energy, Eq.~(\ref{eq:selfenergy}): emission of a magnon pair and its later reabsorption; 
the double line is the dressed singlet propagator.}
\end{figure*}

In this Appendix we give the explicit form of the Hamiltonian solved within the SCBA, Eq.~(\ref{eq:hamil}), 
and show how it follows from the two-orbital Hubbard model, Eq.~(\ref{hhub}). The derivation proceeds in three steps: (i) a strong-coupling expansion of $H_{\rm Hub}$, which yields a 
$t$--$J$-like model for triplets and singlets; (ii) the completion of this model with the smaller 
first-neighbor terms and with the spin Hamiltonian whose parameters are taken from neutron scattering experiments; 
and (iii) the spin-wave representation of the zigzag background, which separates the singlet hopping into free and magnon-assisted terms. The processes retained in the final Hamiltonian are illustrated in
Fig.~\ref{fig:cartoon}. 

\subsection{$t$--$J$-like model}

Retaining the dominant third-NN hoppings of Eq.~(\ref{hhub}) and using second-order perturbation theory in 
$H_{\rm hop}$, we obtain an effective Hamiltonian that describes the antiferromagnetic exchange between third-NN
spins $S=1$ and the motion of the singlets $b$ and $c$ in the magnetic background~\cite{Hamad2024}. 
Here $\bm\delta_j$ ($j=1,2,3$) are the three third-NN vectors of Fig.~\ref{fig:cartoon}(a), with 
$\bm\delta_1\parallel b$:
\begin{widetext}
\begin{eqnarray}
H_{\text{t-J}} &=&\sum\limits_{i}\left[ (E_{b}-E_{t})b_{i}^{\dagger}b_{i}+(E_{c}-E_{t})c_{i}^{\dagger }c_{i}\right]
+\sum\limits_{i\in A}\Big[J^{(\delta_1)}_{3}\mathbf{S}_{i+\delta _{1}}\cdot \mathbf{S}_{i}
+J^{(\delta_{2,3})}_{3}\sum\limits_{\delta _{j}\neq \delta _{1}}\mathbf{S}_{i+\delta_{j}}\cdot \mathbf{S}_{i}\Big] \notag \\
& + & h_{cc}^{(\delta_1)}\sum\limits_{i\in A}\sum\limits_{m}\left( c_{i+\delta _{1}}^{\dagger}t_{im}^{\dagger }t_{i+\delta _{1}m}c_{i}+\text{H.c.}\right)
+h_{cc}^{(\delta_{2,3})}\sum\limits_{i\in A}\sum\limits_{\delta _{j}\neq \delta _{1}}\sum\limits_{m}\left(
c_{i+\delta _{j}}^{\dagger }t_{im}^{\dagger }t_{i+\delta _{j}m}c_{i}+\text{H.c.}\right) \notag \\
& + & h_{bc}^{(\delta_{2,3})}\sum\limits_{i\in A}\sum\limits_{\delta _{j}\neq \delta_{1}}(-1)^{j}\sum\limits_{m}\left(c_{i+\delta _{j}}^{\dagger }t_{im}^{\dagger}t_{i+\delta _{j}m}b_{i}
+b_{i+\delta _{j}}^{\dagger }t_{im}^{\dagger }t_{i+\delta_{j}m}c_{i}+\text{H.c.}\right),
\label{HtJ}
\end{eqnarray}
\end{widetext}
where the bosonic operators $b_{i}$, $c_{i}$ and $t_{im}$ ($m=S^z$) represent
the two-particle states of Eq.~(\ref{ee}), and
\begin{eqnarray}
&&J^{(\delta_1)}_{3} =\frac{t_{1}^{2}}{U+J_H},\quad
J^{(\delta_{2,3})}_{3}=\frac{t_{2}^{2}+2t_{3}^{2}}{U+J_H},  \notag \\
&&h_{cc}^{(\delta_1)}=\frac{t_{1}^{2}}{U-J_H},\quad
h_{cc}^{(\delta_{2,3})} =\frac{t_{2}^{2}+t_{3}^{2}}{U-J_H},  \notag \\
&&h_{bc}^{(\delta_{2,3})}
=\frac{t_{2}t_{3}}{2}\left( \frac{1}{U-J_H}+\frac{1}{U^{\prime }+J^{\prime}}\right).
\label{param}
\end{eqnarray}
 For $U=4$~eV and $J_H=0.75$~eV we
obtain $J^{(\delta_1)}_{3}=10.6$~meV, $J^{(\delta_{2,3})}_{3}=10.3$~meV,
$h_{cc}^{(\delta_1)}=15.5$~meV, $h_{cc}^{(\delta_{2,3})}=11.7$~meV, and
$h_{bc}^{(\delta_{2,3})}=5.3$~meV. 
The hopping terms of Eq.~(\ref{HtJ}) have a simple meaning, which is the key to
the singlet dynamics: the operator
$c^\dagger_{i+\delta}\,t^\dagger_{im}\,t_{i+\delta,m}\,c_i$ interchanges
the singlet at site $i$ with the triplet at site $i+\delta$. The singlet moves
to $i+\delta$, while the triplet moves to $i$ keeping its spin projection $m$
[Figs.~\ref{fig:cartoon}(b) and (c)]. This is the elementary step also
considered in Ref.~\cite{He2024}.  Notice also that
there is no $b$--$b$ term at this order: singlet $b$ can move only by
converting into singlet $c$, on the $\bm\delta_2$ and $\bm\delta_3$ bonds.

\subsection{Complete effective model}

The model actually solved~\cite{Hamad2024} differs from Eq.~(\ref{HtJ}) in two respects.

(i) {\it Magnetic part.} The Heisenberg terms of Eq.~(\ref{HtJ}) are replaced
by the full spin Hamiltonian $H_J$ of Sec.~\ref{sec:SCBA}, with first-, second-
and third-NN XXZ exchange and single-ion anisotropies, and the parameters of
Ref.~\cite{Scheie2023} ($J_1=-2.7$, $J_2=-0.42$, $J_3=13.9$, $D_1=-0.01$,
$D_2=0.21$~meV, $\Delta=1$). This guarantees that the magnons coupled to the
singlets are those measured by inelastic neutron scattering. The perturbative
values $J_3\simeq10.3$--$10.6$~meV obtained above are consistent with this
choice.

(ii) {\it First-neighbor hoppings.} Equation~(\ref{HtJ}) splits the lattice
into four decoupled honeycomb sublattices. They are connected by the smaller
first-NN hoppings of the Wannier model, whose contribution to the singlet
hopping, estimated with the same perturbation theory, is about 10\% of the
third-NN values. Second-NN hoppings are negligible and are set to zero.

For a generic bond $\langle ij\rangle$ the singlet part of the resulting model reads
\begin{eqnarray}
H_s &=& \sum_i\big[\varepsilon_b\, b^\dagger_i b_i+\varepsilon_c\, c^\dagger_i c_i\big]
\notag\\
&+&\sum_{\langle ij\rangle}\sum_{\alpha\beta} t^{(ij)}_{\alpha\beta}\sum_m
\big(\alpha^\dagger_j t^\dagger_{im} t_{jm} \beta_i+\text{H.c.}\big),
\label{Hs}
\end{eqnarray}
with $\alpha,\beta=b,c$, $\varepsilon_\alpha=E_\alpha-E_t$, and symmetric amplitudes
$t^{(ij)}_{\alpha\beta}=t^{(ij)}_{\beta\alpha}$. On third-NN bonds,
$t_{cc}=h_{cc}$ and $t_{bc}=(-1)^j h_{bc}$ as in Eq.~(\ref{HtJ}). The values used
in all the calculations of this work are listed in Table~\ref{tab:hop}.

\begin{table}[t]
\caption{\label{tab:hop} Singlet hopping amplitudes (meV), classified according
to the alignment of the two spins in the zigzag order. $t_{bb}=0$ on all bonds.
The first-NN amplitudes are 10\% of those of the antiparallel third-NN bond.}
\begin{ruledtabular}
\begin{tabular}{@{}lcccc@{}}
bond & zigzag & $t_{cc}$ & $t_{bc}$ \\
\hline
third NN, $\bm\delta_1$ ($\parallel b$)    & AF & 15.5 & 0 \\
third NN, $\bm\delta_2,\bm\delta_3$        & AF & 11.7 & $\pm5.3$ \\
first NN, across chains ($\parallel b$)    & AF & 1.55 & 0 \\
first NN, along the chain                  & FM & 1.17 & $\pm0.5$ \\
second NN                                  & 2 FM, 4 AF & 0 & 0 \\
\end{tabular}
\end{ruledtabular}
\end{table}

\subsection{Spin waves of the zigzag background}

The spin Hamiltonian $H_J$ is treated in linear spin-wave theory (Sec.~\ref{sec:SCBA}). We take
the unit cell with sublattice $A$ at the origin and $B$ at $(0,a)$, $a=1/\sqrt3$, lengths in units of
the lattice constant, and $a_{i\xi}=N^{-1/2}\sum_{\bm k}e^{-i\bm k\cdot\bm r_i}a_{\bm k\xi}$, with
$N$ the number of cells and $\bm r_i$ the position of the cell (the same convention is used for the singlets). After the local
rotation that maps the zigzag order onto a ferromagnetic one, the dynamical matrix in the basis
$(a_{\bm kA},a_{\bm kB},a^\dagger_{-\bm kA},a^\dagger_{-\bm kB})$ is
\begin{equation}
\mathcal M_{\bm k}=\begin{pmatrix}
\epsilon_{\bm k}&g_{\bm k}&h_{\bm k}&\gamma_{\bm k}\\
g^*_{\bm k}&\epsilon_{\bm k}&\gamma^*_{\bm k}&h_{\bm k}\\
h_{\bm k}&\gamma_{\bm k}&\epsilon_{\bm k}&g_{\bm k}\\
\gamma^*_{\bm k}&h_{\bm k}&g^*_{\bm k}&\epsilon_{\bm k}
\end{pmatrix},
\label{MkA}
\end{equation}
with
\begin{equation}
\begin{aligned}
\epsilon_{\bm k}&=S(-J_1+2J_2+3J_3+D_2-2D_1)\\
&\quad+SJ_2\big[(1+\Delta)\cos k_x-(1-\Delta)\,2\cos\tfrac{k_x}2\cos\tfrac{\sqrt3k_y}2\big],\\
h_{\bm k}&=SJ_2\big[(1-\Delta)\cos k_x-(1+\Delta)\,2\cos\tfrac{k_x}2\cos\tfrac{\sqrt3k_y}2\big]\\
&\quad-SD_2,\\
\gamma_{\bm k}&=-\tfrac{SJ_1}2(1+\Delta)+SJ_1(1-\Delta)\,e^{i\sqrt3k_y/2}\cos\tfrac{k_x}2\\
&\quad-\tfrac{SJ_3}2(1+\Delta)\big(2\cos k_x+e^{i\sqrt3k_y}\big),\\
g_{\bm k}&=-\tfrac{SJ_1}2(1-\Delta)+SJ_1(1+\Delta)\,e^{i\sqrt3k_y/2}\cos\tfrac{k_x}2\\
&\quad-\tfrac{SJ_3}2(1-\Delta)\big(2\cos k_x+e^{i\sqrt3k_y}\big).
\end{aligned}
\label{MelemA}
\end{equation}
 The magnon energies $\omega_{\bm k\eta}$ and the
Bogoliubov coefficients,
$a_{\bm k\xi}=\sum_\eta\big[U_{\xi\eta}(\bm k)\alpha_{\bm k\eta}+V_{\xi\eta}(\bm k)\alpha^\dagger_{-\bm k\eta}\big]$,
follow from its paraunitary diagonalization~\cite{COLPA1978}. The four branches of the magnetic
Brillouin zone are obtained by folding $\omega_{\bm k\eta}$ and $\omega_{\bm k+\bm Q,\eta}$. 

\subsection{Holstein-Primakov representation and singlet-magnon vertex}

We follow the standard treatment of a hole in an ordered
antiferromagnet~\cite{Kane1989,Hamad2006,Hamad2008}. In the rotated frame the triplet is represented
by Holstein--Primakov bosons, $|\tilde m=1-n\rangle\leftrightarrow|n\rangle$, with $n$ the number of spin deviations ($\tilde m=\pm m$ on the
up/down sublattice), and the site occupied by the singlet is assigned the boson vacuum. The singlet is described by
$h_{i\alpha}$ ($h_{ib}=b_i$, $h_{ic}=c_i$), the real-space counterpart of the operators
$h_{\bm k\xi\alpha}$ of Sec.~\ref{sec:SCBA}.
The hop of Eq.~(\ref{Hs}) then takes the vacuum at $i$ to $|n_i\rangle$ and $|n_j\rangle$ at $j$ to
the vacuum, with $n_i=n_j$ on ferromagnetic (FM) bonds and $n_i+n_j=2$ on antiferromagnetic (AF) ones.
Retaining for the moving triplet only the projections $\tilde m=\pm1$ of the classical order
(i.e., neglecting the channel $n_i=n_j=1$) and expanding to second order in the bosons, we obtain
\begin{eqnarray}
H_t&=&\sum_{\langle ij\rangle}\sum_{\alpha\beta}t^{(ij)}_{\alpha\beta}
\big[h^\dagger_{j\alpha}h_{i\beta}\,\Gamma_{ij}+\text{H.c.}\big],
\label{Ht}\\
\Gamma^{\rm FM}_{ij}&=&1,\qquad
\Gamma^{\rm AF}_{ij}=\tfrac{1}{\sqrt2}\big(a^\dagger_ia^\dagger_i+a_ja_j\big),
\label{Gamma}
\end{eqnarray}
where $i$ ($j$) is the site the singlet leaves (reaches). Across an AF bond a triplet aligned with the
order at $j$ arrives reversed at $i$, leaving a deviation $\delta S^z=2$ ($a^\dagger_ia^\dagger_i$);
conversely, a triplet reversed at $j$ by zero-point fluctuations arrives aligned, and the two
deviations are removed ($a_ja_j$) [Fig.~\ref{fig:cartoon}(b)].

$\Gamma^{\rm FM}$ gives the free hopping $H_t^{(0)}$. $\Gamma^{\rm AF}$, written with the Bogoliubov
operators, gives $H_t^{\rm int}=N^{-1}\sum M^{\xi\xi',\eta\eta'}_{\alpha\beta}(\bm k,\bm q,\bm q')\,h^\dagger_{\bm k'\xi'\alpha}h_{\bm k\xi\beta}\,
\alpha^\dagger_{\bm q\eta}\alpha^\dagger_{\bm q'\eta'}+\text{H.c.}$, with $\bm k'=\bm k-\bm q-\bm q'$,
plus singlet--magnon scattering terms that do not contribute to the SCBA self-energy at $T=0$.
With the normalization of Eq.~(\ref{eq:selfenergy}), the vertex for a singlet that leaves sublattice
$\xi$ and reaches $\xi'$ is
\begin{equation}
   \begin{split}
   M^{\xi\xi',\eta\eta'}_{\alpha\beta}(\mathbf{k},\mathbf{q},\mathbf{q}') ={}& \gamma^{\rm AF}_{\alpha\beta}(\mathbf{k}')\,U^*_{\xi\eta}(\mathbf{q})U^*_{\xi\eta'}(\mathbf{q}') \\
   &+ \gamma^{\rm AF}_{\alpha\beta}(\mathbf{k})\,V_{\xi'\eta}(-\mathbf{q})V_{\xi'\eta'}(-\mathbf{q}'),
   \end{split}
   \label{vertexA}
   \end{equation}
%
where $\gamma^{\rm AF}_{\alpha\beta}(\bm p)=\sum_{\bm d}t^{(\bm d)}_{\alpha\beta}e^{-i\bm p\cdot\bm d}$
runs over the AF bonds $\bm d$ that join sublattice $\xi$ to $\xi'$.

The full Hamiltonian is $H_{\rm eff}=H_J+H_t^{(0)}+H_t^{\rm int}$, Eq.~(\ref{eq:hamil}). Coherent propagation arises from sequences in which a hop that creates two deviations is combined with one that removes deviations already present in the ground state, the diagrammatic counterpart of the string repaired by spin exchange in Fig.~5 of Ref.~\cite{He2024}. 

\makeatletter\def\normalcolor{\color{black}}\makeatother\color{black}

\bibliography{references}

@article{Zhang1988,
  title = {{Effective {Hamiltonian} for the superconducting Cu oxides}},
  author = {Zhang, F. C. and Rice, T. M.},
  journal = {Phys. Rev. B},
  volume = {37},
  issue = {7},
  pages = {3759--3761},
  numpages = {0},
  year = {1988},
  month = {Mar},
  publisher = {American Physical Society},
  doi = {10.1103/PhysRevB.37.3759},
  url = {https://link.aps.org/doi/10.1103/PhysRevB.37.3759}
}

@article{Hamad2024,
  title = {Singlet Polaron Theory of Low-Energy Optical Excitations in {${\mathrm{NiPS}}_{3}$}},
  author = {Hamad, I. J. and Helman, C. S. and Manuel, L. O. and Feiguin, A. E. and Aligia, A. A.},
  journal = {Phys. Rev. Lett.},
  volume = {133},
  issue = {14},
  pages = {146502},
  numpages = {6},
  year = {2024},
  month = {Sep},
  publisher = {American Physical Society},
  doi = {10.1103/PhysRevLett.133.146502},
  url = {https://link.aps.org/doi/10.1103/PhysRevLett.133.146502}
}

@article{Aligia2004,
  title = {{Magnetic and orbital ordering of ${\mathrm{RuO}}_{2}$ planes in ${\mathrm{RuSr}}_{2}(\mathrm{Eu},\mathrm{Gd}){\mathrm{Cu}}_{2}{\mathrm{O}}_{8}$}},
  author = {Aligia, A. A. and Gusm\~ao, M. A.},
  journal = {Phys. Rev. B},
  volume = {70},
  issue = {5},
  pages = {054403},
  numpages = {7},
  year = {2004},
  month = {Aug},
  publisher = {American Physical Society},
  doi = {10.1103/PhysRevB.70.054403},
  url = {https://link.aps.org/doi/10.1103/PhysRevB.70.054403}
}

@article{Scheie2023,
  title = {{Spin wave {Hamiltonian} and anomalous scattering in ${\mathrm{NiPS}}_{3}$}},
  author = {Scheie, A. and Park, Pyeongjae and Villanova, J. W. and Granroth, G. E. and Sarkis, C. L. and Zhang, Hao and Stone, M. B. and Park, Je-Geun and Okamoto, S. and Berlijn, T. and Tennant, D. A.},
  journal = {Phys. Rev. B},
  volume = {108},
  issue = {10},
  pages = {104402},
  numpages = {14},
  year = {2023},
  month = {Sep},
  publisher = {American Physical Society},
  doi = {10.1103/PhysRevB.108.104402},
  url = {https://link.aps.org/doi/10.1103/PhysRevB.108.104402}
}

@misc{plumley2023,
      title={{3D {Heisenberg} universality in the van der {Waals} antiferromagnet {${\mathrm{NiPS}}_{3}$}}},
      author={Rajan Plumley and Sougata Mardanya and Cheng Peng and Johannes Nokelainen and Tadesse Assefa and Lingjia Shen and Nicholas Burdet and Zach Porter and Alexander Petsch and Aidan Israelski and Hongwei Chen and Jun Sik Lee and Sophie Morley and Sujoy Roy and Gilberto Fabbris and Elizabeth Blackburn and Adrian Feiguin and Arun Bansil and Wei-Sheng Lee and Aaron Lindenberg and Sugata Chowdhury and Mike Dunne and Joshua J. Turner},
      year={2023},
      eprint={2310.07948},
      archivePrefix={arXiv},
      primaryClass={cond-mat.str-el}
}

@misc{DiScala2023,
      title={{Dimensionality dependent electronic structure of the exfoliated van der Waals antiferromagnet ${\mathrm{NiPS}}_{3}$}},
      author={M. F. DiScala and D. Staros and A. de la Torre and A. Lopez and D. Wong and C. Schulz and M. Bartkowiak and B. Rubenstein and K. W. Plumb},
      year={2023},
      eprint={2302.07910},
      archivePrefix={arXiv},
      primaryClass={cond-mat.str-el}
}

@article{Kim2019,
	author = {Kim, Kangwon and Lim, Soo Yeon and Lee, Jae-Ung and Lee, Sungmin and Kim, Tae Yun and Park, Kisoo and Jeon, Gun Sang and Park, Cheol-Hwan and Park, Je-Geun and Cheong, Hyeonsik},
	date = {2019/01/21},
	doi = {10.1038/s41467-018-08284-6},
	isbn = {2041-1723},
	journal = {Nature Communications},
	number = {1},
	pages = {345},
	title = {{Suppression of magnetic ordering in XXZ-type antiferromagnetic monolayer ${\mathrm{NiPS}}_{3}$}},
	url = {https://doi.org/10.1038/s41467-018-08284-6},
	volume = {10},
	year = {2019}
}

@article{Kuo2016,
	author = {Kuo, Cheng-Tai and Neumann, Michael and Balamurugan, Karuppannan and Park, Hyun Ju and Kang, Soonmin and Shiu, Hung Wei and Kang, Jin Hyoun and Hong, Byung Hee and Han, Moonsup and Noh, Tae Won and Park, Je-Geun},
	date = {2016/02/15},
	doi = {10.1038/srep20904},
	isbn = {2045-2322},
	journal = {Scientific Reports},
	number = {1},
	pages = {20904},
	title = {{Exfoliation and Raman Spectroscopic Fingerprint of Few-Layer {${\mathrm{NiPS}}_{3}$} Van der Waals Crystals}},
	url = {https://doi.org/10.1038/srep20904},
	volume = {6},
	year = {2016},
}

@article{Olsen2021,
archivePrefix = {arXiv},
arxivId = {2103.04474},
author = {Olsen, Thomas},
doi = {10.1088/1361-6463/ac000e},
eprint = {2103.04474},
issn = {13616463},
journal = {J. Phys. D. Appl. Phys.},
number = {31},
title = {{Magnetic anisotropy and exchange interactions of two-dimensional {${\mathrm{FePS}}_{3}$}. {${\mathrm{NiPS}}_{3}$} and {${\mathrm{MnPS}}_{3}$}  from first principles calculations}},
volume = {54},
year = {2021}
}

@article{Huang2026,
  title = {Magnetic-Field-Tunable Repulsive Exciton-Exciton Interaction in the van der Waals Antiferromagnet ${\mathrm{NiPS}}_{3}$},
  author = {Huang, Kaiyang and Park, Jaena and Yang, Zhuo and Park, Je-Geun and Miyata, Atsuhiko and Kohama, Yoshimitsu and Matsuda, Yasuhiro H.},
  journal = {Phys. Rev. Lett.},
  volume = {137},
  issue = {7},
  pages = {076903},
  numpages = {7},
  year = {2026},
  month = {Aug},
  publisher = {American Physical Society},
  doi = {10.1103/jl7j-g835},
  url = {https://link.aps.org/doi/10.1103/jl7j-g835}
}

@article{Wildes2022,
author = {Wildes, A. R. and Stewart, J. R. and Le, M. D. and Ewings, R. A. and Rule, K. C. and Deng, G. and Anand, K.},
doi = {10.1103/PhysRevB.106.174422},
issn = {24699969},
journal = {Phys. Rev. B},
number = {17},
pages = {1--11},
title = {{Magnetic dynamics of {${\mathrm{NiPS}}_{3}$}}},
volume = {106},
year = {2022}
}

@article{Afanasiev2021,
archivePrefix = {arXiv},
arxivId = {2010.05062},
author = {Afanasiev, Dmytro and Hortensius, Jorrit R. and Matthiesen, Mattias and Ma{\~{n}}as-Valero, Samuel and {\v{S}}i{\v{s}}kins, Makars and Lee, Martin and Lesne, Edouard and van der Zant, Herre S.J. and Steeneken, Peter G. and Ivanov, Boris A. and Coronado, Eugenio and Caviglia, Andrea D.},
doi = {10.1126/sciadv.abf3096},
eprint = {2010.05062},
issn = {23752548},
journal = {Sci. Adv.},
number = {23},
pages = {1--8},
title = {{Controlling the anisotropy of a van der Waals antiferromagnet with light}},
volume = {7},
year = {2021}
}

@misc{Mi2022,
archivePrefix = {arXiv},
arxivId = {2112.02544},
author = {Mi, Mengjuan and Zheng, Xingwen and Wang, Shilei and Zhou, Yang and Yu, Lixuan and Xiao, Han and Song, Houning and Shen, Bing and Li, Fangsen and Bai, Lihui and Chen, Yanxue and Wang, Shanpeng and Liu, Xiaohui and Wang, Yilin},
booktitle = {Adv. Funct. Mater.},
doi = {10.1002/adfm.202112750},
eprint = {2112.02544},
issn = {16163028},
number = {29},
title = {{Variation between Antiferromagnetism and Ferrimagnetism in NiPS3 by Electron Doping}},
volume = {32},
year = {2022}
}

@article{Lee2016,
archivePrefix = {arXiv},
arxivId = {1608.04169},
author = {Lee, Jae Ung and Lee, Sungmin and Ryoo, Ji Hoon and Kang, Soonmin and Kim, Tae Yun and Kim, Pilkwang and Park, Cheol Hwan and Park, Je Geun and Cheong, Hyeonsik},
doi = {10.1021/acs.nanolett.6b03052},
eprint = {1608.04169},
issn = {15306992},
journal = {Nano Lett.},
number = {12},
pages = {7433--7438},
title = {{Ising-Type Magnetic Ordering in Atomically Thin {${\mathrm{FePS}}_{3}$}}},
volume = {16},
year = {2016}
}

@article{Gibertini2019,
author = {Gibertini, M. and Koperski, M. and Morpurgo, A. F. and Novoselov, K. S.},
doi = {10.1038/s41565-019-0438-6},
issn = {17483395},
journal = {Nat. Nanotechnol.},
number = {5},
pages = {408--419},
pmid = {31065072},
publisher = {Springer US},
title = {{Magnetic 2D materials and heterostructures}},
url = {http://dx.doi.org/10.1038/s41565-019-0438-6},
volume = {14},
year = {2019}
}

@article{Burch2018,
author = {Burch, Kenneth S. and Mandrus, David and Park, Je Geun},
doi = {10.1038/s41586-018-0631-z},
issn = {14764687},
journal = {Nature},
number = {7729},
pages = {47--52},
pmid = {30382199},
publisher = {Springer US},
title = {{Magnetism in two-dimensional van der Waals materials}},
url = {http://dx.doi.org/10.1038/s41586-018-0631-z},
volume = {563},
year = {2018}
}

@article{Kang2020,
author = {Kang, Soonmin and Kim, Kangwon and Kim, Beom Hyun and Kim, Jonghyeon and Sim, Kyung Ik and Lee, Jae Ung and Lee, Sungmin and Park, Kisoo and Yun, Seokhwan and Kim, Taehun and Nag, Abhishek and Walters, Andrew and Garcia-Fernandez, Mirian and Li, Jiemin and Chapon, Laurent and Zhou, Ke Jin and Son, Young Woo and Kim, Jae Hoon and Cheong, Hyeonsik and Park, Je Geun},
doi = {10.1038/s41586-020-2520-5},
isbn = {4158602025},
issn = {14764687},
journal = {Nature},
number = {7818},
pages = {785--789},
pmid = {32690938},
publisher = {Springer US},
title = {{Coherent many-body exciton in van der Waals antiferromagnet {${\mathrm{NiPS}}_{3}$}}},
url = {http://dx.doi.org/10.1038/s41586-020-2520-5},
volume = {583},
year = {2020}
}

@article{Dirnberger2022,
author = {Dirnberger, Florian and Bushati, Rezlind and Datta, Biswajit and Kumar, Ajesh and MacDonald, Allan H. and Baldini, Edoardo and Menon, Vinod M.},
doi = {10.1038/s41565-022-01204-2},
isbn = {4156502201},
issn = {17483395},
journal = {Nat. Nanotechnol.},
number = {10},
pages = {1060--1064},
pmid = {36097046},
publisher = {Springer US},
title = {{Spin-correlated exciton–polaritons in a van der Waals magnet}},
volume = {17},
year = {2022}
}

@article{Belvin2021,
archivePrefix = {arXiv},
arxivId = {2106.08355},
author = {Belvin, Carina A. and Baldini, Edoardo and Ozel, Ilkem Ozge and Mao, Dan and Po, Hoi Chun and Allington, Clifford J. and Son, Suhan and Kim, Beom Hyun and Kim, Jonghyeon and Hwang, Inho and Kim, Jae Hoon and Park, Je Geun and Senthil, T. and Gedik, Nuh},
doi = {10.1038/s41467-021-25164-8},
eprint = {2106.08355},
issn = {20411723},
journal = {Nat. Commun.},
number = {1},
pages = {6--12},
pmid = {34376692},
publisher = {Springer US},
title = {{Exciton-driven antiferromagnetic metal in a correlated van der Waals insulator}},
url = {http://dx.doi.org/10.1038/s41467-021-25164-8},
volume = {12},
year = {2021}
}

@article{Kim2018,
archivePrefix = {arXiv},
arxivId = {1706.06259},
author = {Kim, So Yeun and Kim, Tae Yun and Sandilands, Luke J. and Sinn, Soobin and Lee, Min Cheol and Son, Jaeseok and Lee, Sungmin and Choi, Ki Young and Kim, Wondong and Park, Byeong Gyu and Jeon, C. and Kim, Hyeong Do and Park, Cheol Hwan and Park, Je Geun and Moon, S. J. and Noh, T. W.},
doi = {10.1103/PhysRevLett.120.136402},
eprint = {1706.06259},
issn = {10797114},
journal = {Phys. Rev. Lett.},
number = {13},
pages = {136402},
pmid = {29694193},
publisher = {American Physical Society},
title = {{Charge-Spin Correlation in van der Waals Antiferromagnet ${\mathrm{NiPS}}_{3}$}},
url = {https://doi.org/10.1103/PhysRevLett.120.136402},
volume = {120},
year = {2018}
}

@article{Wildes2015,
  title = {Magnetic structure of the quasi-two-dimensional antiferromagnet {${\text{NiPS}}_{3}$}},
  author = {Wildes, A. R. and Simonet, V. and Ressouche, E. and McIntyre, G. J. and Avdeev, M. and Suard, E. and Kimber, S. A. J. and Lan\ifmmode \mbox{\c{c}}\else \c{c}\fi{}on, D. and Pepe, G. and Moubaraki, B. and Hicks, T. J.},
  journal = {Phys. Rev. B},
  volume = {92},
  issue = {22},
  pages = {224408},
  numpages = {11},
  year = {2015},
  month = {Dec},
  publisher = {American Physical Society},
  doi = {10.1103/PhysRevB.92.224408},
  url = {https://link.aps.org/doi/10.1103/PhysRevB.92.224408}
}

@article{Joy1992,
  title = {Magnetism in the layered transition-metal thiophosphates {M${\mathrm{PS}}_{3}$ (M=Mn, Fe, and Ni)}},
  author = {Joy, P. A. and Vasudevan, S.},
  journal = {Phys. Rev. B},
  volume = {46},
  issue = {9},
  pages = {5425--5433},
  numpages = {0},
  year = {1992},
  month = {Sep},
  publisher = {American Physical Society},
  doi = {10.1103/PhysRevB.46.5425},
  url = {https://link.aps.org/doi/10.1103/PhysRevB.46.5425}
}

@article{Toth2015,
doi = {10.1088/0953-8984/27/16/166002},
url = {https://dx.doi.org/10.1088/0953-8984/27/16/166002},
year = {2015},
month = {mar},
publisher = {IOP Publishing},
volume = {27},
number = {16},
pages = {166002},
author = {S Toth and B Lake},
title = {Linear spin wave theory for single-Q incommensurate magnetic structures},
journal = {Journal of Physics: Condensed Matter}
}

@article{COLPA1978,
title = {Diagonalization of the quadratic boson {Hamiltonian}},
journal = {Physica A: Statistical Mechanics and its Applications},
volume = {93},
number = {3},
pages = {327-353},
year = {1978},
issn = {0378-4371},
doi = {https://doi.org/10.1016/0378-4371(78)90160-7},
url = {https://www.sciencedirect.com/science/article/pii/0378437178901607},
author = {J.H.P. Colpa}
}

@article{Hamad2008,
  title = {Spin polaron in the ${J}_{1}\text{\ensuremath{-}}{J}_{2}$ {Heisenberg} model},
  author = {Hamad, I. J. and Trumper, A. E. and Feiguin, A. E. and Manuel, L. O.},
  journal = {Phys. Rev. B},
  volume = {77},
  issue = {1},
  pages = {014410},
  numpages = {9},
  year = {2008},
  month = {Jan},
  publisher = {American Physical Society},
  doi = {10.1103/PhysRevB.77.014410},
  url = {https://link.aps.org/doi/10.1103/PhysRevB.77.014410}
}

@article{Kane1989,
  title = {Motion of a single hole in a quantum antiferromagnet},
  author = {Kane, C. L. and Lee, P. A. and Read, N.},
  journal = {Phys. Rev. B},
  volume = {39},
  issue = {10},
  pages = {6880--6897},
  numpages = {0},
  year = {1989},
  month = {Apr},
  publisher = {American Physical Society},
  doi = {10.1103/PhysRevB.39.6880},
  url = {https://link.aps.org/doi/10.1103/PhysRevB.39.6880}
}

@article{Jana2023,
  title = {{Magnon gap excitations and spin-entangled optical transition in the van der Waals antiferromagnet ${\text{NiPS}}_{3}$}},
  author = {Jana, Dipankar and Kapuscinski, P. and Mohelsky, I. and Vaclavkova, D. and Breslavetz, I. and Orlita, M. and Faugeras, C. and Potemski, M.},
  journal = {Phys. Rev. B},
  volume = {108},
  issue = {11},
  pages = {115149},
  numpages = {8},
  year = {2023},
  month = {Sep},
  publisher = {American Physical Society},
  doi = {10.1103/PhysRevB.108.115149},
  url = {https://link.aps.org/doi/10.1103/PhysRevB.108.115149}
}

@article{Klaproth2023,
  title = {{Origin of the Magnetic Exciton in the van der Waals Antiferromagnet ${\mathrm{NiPS}}_{3}$}},
  author = {Klaproth, T. and Aswartham, S. and Shemerliuk, Y. and Selter, S. and Janson, O. and van den Brink, J. and B\"uchner, B. and Knupfer, M. and Pazek, S. and Mikhailova, D. and Efimenko, A. and Hayn, R. and Savoyant, A. and Gubanov, V. and Koitzsch, A.},
  journal = {Phys. Rev. Lett.},
  volume = {131},
  issue = {25},
  pages = {256504},
  numpages = {7},
  year = {2023},
  month = {Dec},
  publisher = {American Physical Society},
  doi = {10.1103/PhysRevLett.131.256504},
  url = {https://link.aps.org/doi/10.1103/PhysRevLett.131.256504}
}

@article{Kim2023,
title={{Rapid Suppression of Quantum Many-Body Magnetic Exciton in Doped van der Waals Antiferromagnet {${\mathrm{(Ni,Cd)PS}}_{3}$}}},
author = {Kim, Junghyun and Na, Woongki and Kim, Jonghyeon and Park, Pyeongjae and Zhang, Kaixuan and Hwang, Inho and Son, Young-Woo and Kim, Jae Hoon and Cheong, Hyeonsik and Park, Je-Geun},
year = {2023},
journal = {Nano Letters},
volume = {23},
pages = {10189},
doi = {10.1021/acs.nanolett.3c02677},
url = {https://doi.org/10.1021/acs.nanolett.3c02677}
}

@Article{Hwangbo2021,
author={Hwangbo, Kyle
and Zhang, Qi
and Jiang, Qianni
and Wang, Yong
and Fonseca, Jordan
and Wang, Chong
and Diederich, Geoffrey M.
and Gamelin, Daniel R.
and Xiao, Di
and Chu, Jiun-Haw
and Yao, Wang
and Xu, Xiaodong},
title={{Highly anisotropic excitons and multiple phonon bound states in a van der Waals antiferromagnetic insulator}},
journal={Nature Nanotechnology},
year={2021},
month={Jun},
day={01},
volume={16},
pages={655-660}
}

@Article{Kim2021,
author={Kim, Tae Yun
and Park, Cheol-Hwan},
title={Magnetic Anisotropy and Magnetic Ordering of Transition-Metal Phosphorus Trisulfides},
journal={Nano Letters},
year={2021},
month={Dec},
day={08},
publisher={American Chemical Society},
volume={21},
number={23},
pages={10114-10121},
issn={1530-6984},
doi={10.1021/acs.nanolett.1c03992},
url={https://doi.org/10.1021/acs.nanolett.1c03992}
}

@Article{Wang2021,
author={Wang, Xingzhi
and Cao, Jun
and Lu, Zhengguang
and Cohen, Arielle
and Kitadai, Hikari
and Li, Tianshu
and Tan, Qishuo
and Wilson, Matthew
and Lui, Chun Hung
and Smirnov, Dmitry
and Sharifzadeh, Sahar
and Ling, Xi},
title={{Spin-induced linear polarization of photoluminescence in antiferromagnetic van der Waals crystals}},
journal={Nature Materials},
year={2021},
month={Jul},
day={01},
volume={20},
number={7},
pages={964-970}
}

@Article{Ho2021,
author={Ho, Ching-Hwa
and Hsu, Tien-Yao
and Muhimmah, Luthviyah Choirotul},
title={The band-edge excitons observed in few-layer {${\mathrm{(Ni,Cd)PS}}_{3}$}},
journal={npj 2D Materials and Applications},
year={2021},
month={Jan},
day={05},
volume={5},
number={1},
pages={8},
issn={2397-7132},
doi={10.1038/s41699-020-00188-8},
url={https://doi.org/10.1038/s41699-020-00188-8}
}

@article{Mehlawat2022,
  title = {{Low-energy excitations and magnetic anisotropy of the layered van der Waals antiferromagnet ${\mathrm{Ni}}_{2}{\mathrm{P}}_{2}{\mathrm{S}}_{6}$}},
  author = {Mehlawat, K. and Alfonsov, A. and Selter, S. and Shemerliuk, Y. and Aswartham, S. and B\"uchner, B. and Kataev, V.},
  journal = {Phys. Rev. B},
  volume = {105},
  issue = {21},
  pages = {214427},
  numpages = {7},
  year = {2022},
  month = {Jun},
  publisher = {American Physical Society},
  doi = {10.1103/PhysRevB.105.214427},
  url = {https://link.aps.org/doi/10.1103/PhysRevB.105.214427}
}

@article{KimKim2023,
doi = {10.1088/1367-2630/aced64},
url = {https://dx.doi.org/10.1088/1367-2630/aced64},
year = {2023},
month = {aug},
publisher = {IOP Publishing},
volume = {25},
number = {8},
pages = {083029},
author = {Hyeon Jung Kim and Ki-Seok Kim},
title = {Microscopic origin of local electric polarization in {${\mathrm{NiPS}}_{3}$}},
journal = {New Journal of Physics}
}

@article{Occhialini2024,
  title = {{Nature of Excitons and Their Ligand-Mediated Delocalization in Nickel Dihalide Charge-Transfer Insulators}},
  author = {Occhialini, Connor A. and Tseng, Yi and Elnaggar, Hebatalla and Song, Qian and Blei, Mark and Tongay, Seth Ariel and Bisogni, Valentina and de Groot, Frank M. F. and Pelliciari, Jonathan and Comin, Riccardo},
  journal = {Phys. Rev. X},
  volume = {14},
  issue = {3},
  pages = {031007},
  numpages = {18},
  year = {2024},
  month = {Jul},
  publisher = {American Physical Society},
  doi = {10.1103/PhysRevX.14.031007},
  url = {https://link.aps.org/doi/10.1103/PhysRevX.14.031007}
}

@article{Allington2025,
  title = {{Distinct Optical Excitation Mechanisms of a Coherent Magnon in a van der Waals Antiferromagnet}},
  author = {Allington, Clifford J. and Belvin, Carina A. and Seifert, Urban F. P. and Ye, Mengxing and Tai, Tommy and Baldini, Edoardo and Son, Suhan and Kim, Junghyun and Park, Jaena and Park, Je-Geun and Balents, Leon and Gedik, Nuh},
  journal = {Phys. Rev. Lett.},
  volume = {134},
  issue = {6},
  pages = {066903},
  numpages = {6},
  year = {2025},
  month = {Feb},
  publisher = {American Physical Society},
  doi = {10.1103/PhysRevLett.134.066903},
  url = {https://link.aps.org/doi/10.1103/PhysRevLett.134.066903}
}

@article{He2024,
	author = {He, W. and Shen, Y. and Wohlfeld, K. and Sears, J. and Li, J. and Pelliciari, J. and Walicki, M. and Johnston, S. and Baldini, E. and Bisogni, V. and Mitrano, M. and Dean, M. P. M.},
	date = {2024/04/25},
	doi = {10.1038/s41467-024-47852-x},
	id = {He2024},
	isbn = {2041-1723},
	journal = {Nature Communications},
	number = {1},
	pages = {3496},
	title = {{Magnetically propagating Hund's exciton in van der Waals antiferromagnet ${\mathrm{NiPS}}_{3}$}},
	url = {https://doi.org/10.1038/s41467-024-47852-x},
	volume = {15},
	year = {2024}}

@article{Cojo26,
title = {Orbital-driven electronic anisotropy in nickel phosphorous trisulfide},
journal = {Surfaces and Interfaces},
volume = {84},
pages = {108571},
year = {2026},
issn = {2468-0230},
doi = {https://doi.org/10.1016/j.surfin.2026.108571},
url = {https://www.sciencedirect.com/science/article/pii/S2468023026001586},
author = {Iulia Cojocariu and Debasis Dutta and Claudiu Caraiani and Matteo Jugovac and Claudiu Locovei and Tevfik Onur Menteş and Chia-Nung Kuo and Chin Shan Lue and Gianluca D’Olimpio and Victor Kuncser and Amit Agarwal and Andrea Locatelli and Bogdana Borca}
}

@article{Wang25,
  title = {{Correlation between exciton dynamics and spin structure in the van der Waals antiferromagnet ${\mathrm{NiPS}}_{3}$}},
  author = {Wang, Kang and Peng, Yingchen and Huang, Boying and Zhou, Chun and Sun, Qianlu and Tang, Fujie and Xu, Weigao and Du, Kezhao and Wang, Xingzhi and Yang, Ye},
  journal = {Phys. Rev. B},
  volume = {112},
  issue = {22},
  pages = {L220413},
  numpages = {8},
  year = {2025},
  month = {Dec},
  publisher = {American Physical Society},
  doi = {10.1103/nl4q-ghdg},
  url = {https://link.aps.org/doi/10.1103/nl4q-ghdg}
}

@article{Wars25,
  title = {Long-Lived Population Inversion in Resonantly Driven Excitonic Antiferromagnet},
  author = {Warshauer, Jacob A. and Chen, Huyongqing and Bustamante Lopez, Daniel Alejandro and Tan, Qishuo and Tang, Jing and Ling, Xi and Hu, Wanzheng},
  journal = {Phys. Rev. Lett.},
  volume = {134},
  issue = {1},
  pages = {016901},
  numpages = {6},
  year = {2025},
  month = {Jan},
  publisher = {American Physical Society},
  doi = {10.1103/PhysRevLett.134.016901},
  url = {https://link.aps.org/doi/10.1103/PhysRevLett.134.016901}
}

@article{Wars25b,
  title = {{Characteristic exciton energy scales in antiferromagnetic ${\mathrm{NiPS}}_{3}$}},
  author = {Warshauer, Jacob A. and Chen, Huyongqing and Tan, Qishuo and Tang, Jing and Ling, Xi and Hu, Wanzheng},
  journal = {Phys. Rev. B},
  volume = {111},
  issue = {15},
  pages = {155111},
  numpages = {6},
  year = {2025},
  month = {Apr},
  publisher = {American Physical Society},
  doi = {10.1103/PhysRevB.111.155111},
  url = {https://link.aps.org/doi/10.1103/PhysRevB.111.155111}
}

@misc{Rybak26,
      title={{Evidence for Many-Body States in NiPS$_3$ Revealed by Angle-Resolved Photoelectron Spectroscopy}}, 
      author={Miłosz Rybak and Benjamin Pestka and Biplab Bhattacharyya and Jeff Strasdas and Adam K. Budniak and Adi Harchol and Vitaliy Feyer and Iulia Cojocariu and Daniel Baranowski and Efrat Lifshitz and Markus Morgenstern and Magdalena Birowska and Krzysztof Wohlfeld},
      year={2026},
      eprint={2602.03600},
      archivePrefix={arXiv},
      primaryClass={cond-mat.str-el},
      url={https://arxiv.org/abs/2602.03600}, 
}

@article{Wang22,
author = {Xingzhi Wang  and Jun Cao  and Hua Li  and Zhengguang Lu  and Arielle Cohen  and Anubhab Haldar  and Hikari Kitadai  and Qishuo Tan  and Kenneth S. Burch  and Dmitry Smirnov  and Weigao Xu  and Sahar Sharifzadeh  and Liangbo Liang  and Xi Ling },
title = {Electronic Raman scattering in the 2D antiferromagnet NiPS<sub>3</sub>},
journal = {Science Advances},
volume = {8},
number = {2},
pages = {eabl7707},
year = {2022},
doi = {10.1126/sciadv.abl7707},
URL = {https://www.science.org/doi/abs/10.1126/sciadv.abl7707},
eprint = {https://www.science.org/doi/pdf/10.1126/sciadv.abl7707}}

@article{Hamad2007,
title = {Spectroscopic fingerprints of the frustrated magnetic order in Li2VOSiO4: A t–J model study},
journal = {Physica B: Condensed Matter},
volume = {398},
number = {2},
pages = {416-419},
year = {2007},
issn = {0921-4526},
doi = {https://doi.org/10.1016/j.physb.2007.04.051},
url = {https://www.sciencedirect.com/science/article/pii/S0921452607003298},
author = {I.J. Hamad and A.E. Trumper and L.O. Manuel}
}

@article{Hamad2006,
  title = {Hole dynamics in canted antiferromagnets: Coexistence of many-body and free-like excitations},
  author = {Hamad, I. J. and Manuel, L. O. and Martinez, G. and Trumper, A. E.},
  journal = {Phys. Rev. B},
  volume = {74},
  issue = {9},
  pages = {094417},
  numpages = {8},
  year = {2006},
  month = {Sep},
  publisher = {American Physical Society},
  doi = {10.1103/PhysRevB.74.094417},
  url = {https://link.aps.org/doi/10.1103/PhysRevB.74.094417}
}

@misc{helman,
author = {Density Functional Theory calculations},
note={are done with Quantum Espresso code (J.Phys.:Condens.Matter 29, 465901 (2017)) with $a=5.82$\AA, $c=10.08$\AA, cutoff energy of 75 Ry with a k-mesh of 9x9x9. PAW pseudopotential. }
}

@article{Schue26,
  title = {Crystal-Field-Tuned Spin-Flip Luminescence in ${\mathrm{NiPS}}_{3}$},
  author = {Schu\'e, L\'eonard and Pistawala, Nashra and Elnaggar, Hebatalla and Klein, Yannick and Bellin, Christophe and Biscaras, Johan and Sirotti, Fausto and Lassailly, Yves and Cadiz, Fabian and Harnagea, Luminita and Shukla, Abhay},
  journal = {Phys. Rev. Lett.},
  volume = {136},
  issue = {23},
  pages = {236401},
  numpages = {6},
  year = {2026},
  month = {Jun},
  publisher = {American Physical Society},
  doi = {10.1103/7y26-3skd},
  url = {https://link.aps.org/doi/10.1103/7y26-3skd}
}

@article{Kim26,
  title = {{Microscopic evidence for a Zhang-Rice triplet state in the van der Waals antiferromagnet, NiPS${}_{3}$}},
  author = {Kim, Beom Hyun and Lee, Youjin and Hwang, Junik and Kim, Junghyun and Park, Je-Geun and Baek, Seung-Ho},
  journal = {Phys. Rev. Lett.},
  pages = {},
  year = {2026},
  month = {Jun},
  publisher = {American Physical Society},
  doi = {10.1103/hhtt-ffgg},
  url = {https://link.aps.org/doi/10.1103/hhtt-ffgg}
}

@misc{Matsu26,
      title={{Magnetically activated optical visibility of many-body excitons in NiPS$_3$}},
      author={Shuta Matsuura and Shohei Imai and Naoto Tsuji},
      year={2026},
      eprint={2609.12613},
      archivePrefix={arXiv},
      primaryClass={cond-mat.str-el},
      url={https://arxiv.org/abs/2609.12613},
}

\end{document}